\documentclass[12pt]{article}
\usepackage{a4wide}
\usepackage{latexsym,graphicx,epsfig,psfrag,here}
\usepackage{amssymb}
\newcommand{\ba}{\begin{eqnarray}}  
\newcommand{\ea}{\end{eqnarray}}  
\newcommand{\be}{\begin{equation}}  
\newcommand{\ee}{\end{equation}}  
\newcommand{\dis}{\displaystyle}  
\newcommand{\tr}{\mbox{tr}}  
\newcommand{\re}{\mbox{Re}}  
\newcommand{\im}{\mbox{Im}}  
\newcommand{\order}{{\cal O}}  
  
\newcommand{\et}{\!\!\!&&\!\!\!}  
\newcommand{\fr}{\frac}  
\newcommand{\mkd}{m_K^2}

\newcommand{\vpd}{m_{\pi}^2}  
\newcommand{\vpc}{m_{\pi}^4}  
\newcommand{\vkd}{m_K^2}  
\newcommand{\vkc}{m_K^4}

\newcommand{\nn}{\nonumber}

\newcommand{\s}{\sigma}  

\begin{document}
\begin{titlepage}
\vspace{2cm}
\begin{center}

{\large{\bf $K \to 3\pi$ Final State Interactions at NLO in CHPT  
and Cabibbo's Proposal to Measure $a_0-a_2$}}
\vfill
  
{\bf Elvira G\'amiz$^{a)}$\footnote{Present address:
Department of Physics, University of Illinois, Urbana IL 61801, USA.},
 Joaquim Prades$^{b)}$ and 
Ignazio Scimemi$^{c)}$\footnote{Present address:
Center for Theoretical Physics, Laboratory for Nuclear Science, 
 Massachusetts Institute of Technology, Cambridge MA 02139, USA.}}  

\vspace*{1cm}

$^{a)}$ Department of Physics and Astronomy, 
The University of Glasgow \\   
Glasgow G12 8QQ, United Kingdom.\\

$^{b)}$ CAFPE and  Departamento de F\'\i sica Te\'orica 
y del Cosmos, Universidad de Granada,\\ 
Campus de Fuente Nueva, E-18002 Granada, Spain.\\

$^{c)}$  Departament de F\'{\i}sica Te\`orica, IFIC,
 CSIC-Universitat de Val\`encia,\\  
Apt. de Correus 22085, E-46071 Val\`encia, Spain. \\
 
\end{center}
\vfill
\begin{abstract}  
We present the analytical results  
 for the  $K\to3\pi$ final state interaction phases  
at next-to-leading order (NLO) in CHPT.  
We also study the recent Cabibbo's proposal to measure the  
$\pi\pi$ scattering lengths combination $a_0-a_2$ from the    
cusp effect in the $\pi^0\pi^0$  energy spectrum  at threshold   
for $K^+ \to \pi^0 \pi^0 \pi^+$  and $K_L\to \pi^0 \pi^0\pi^0$, and give  
the relevant formulas to describe it  at NLO.
We estimate  the theoretical uncertainty of the $a_0-a_2$ determination  
at NLO in our approach and  obtain that it is not smaller   
than  5 \% if added quadratically  and 7 \% if linearly
for $K^+\to\pi^0\pi^0\pi^+$.  One gets  similar theoretical uncertainties
if the neutral  $K_L\to\pi^0\pi^0\pi^0$ decay  data below threshold
are used instead. For this decay, there are very large  
theoretical uncertainties above threshold 
due to cancellations and  data above threshold 
cannot be used to get the scattering lengths. 
We  do not include isospin corrections  apart of two-pion phase
space factors which are physical. We compare our results for the
cusp effect with  Cabibbo and Isidori's ones.
\end{abstract}
    \date{\today}
  
\end{titlepage}
  
\section{Introduction}  
  
\begin{figure}
\includegraphics[{width=15cm}]{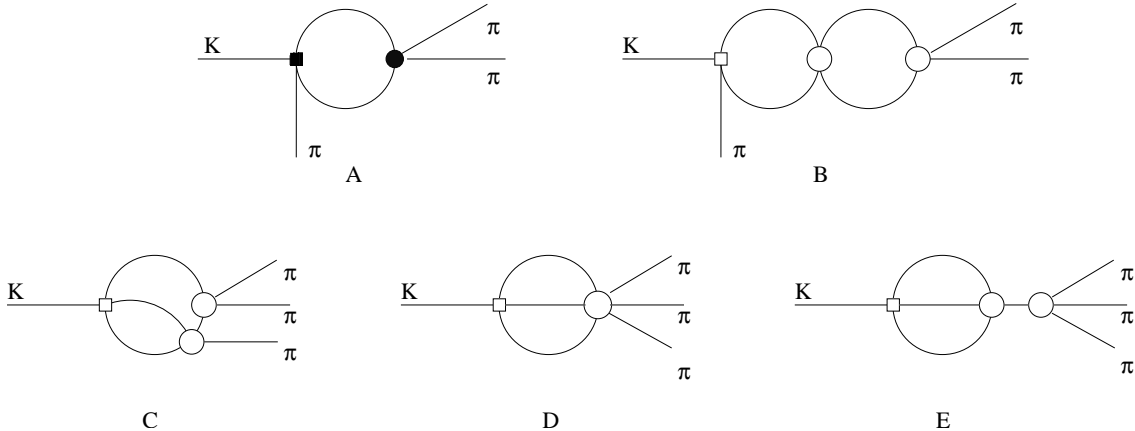}  
        \caption{Topologies contributing to    
FSI in $K \to 3\pi$ decays 
at LO and  NLO. Squares are  weak vertices and   
 circles are  the strong ones. In topology A, solid circle and
solid square vertices include
both tree-  and one-loop level. See text for more details.}  
\label{topologies}
\end{figure}

  Final state interaction (FSI) phases   
at next-to-leading order (NLO) in Chiral Perturbation Theory
\footnote{Some introductory  lectures on CHPT can 
be found in \cite{SOME}   and recent reviews in \cite{reviews}.}
(CHPT)  \cite{WEI79,GL84}
are needed to obtain   the charged CP-violating asymmetries at  NLO  
\cite{GPS03,SGP04,PGS05}. 
 The dominant contribution to these $K^+\to 3\pi$ FSI at NLO  
are from two-pion cuts for topology A in Figure \ref{topologies}. 
They were calculated analytically in \cite{GPS03}.

Though to get the full  
$K\to 3\pi$ amplitudes at order $p^6$  implies a  two-loop calculation,  
 one can get the  FSI phases at NLO  using the optical theorem   
within CHPT   with the advantage that one just needs   
to know  $\pi\pi$ scattering   
and  $K\to 3\pi$ both at ${\cal O}(p^4)$. Notice that  
NLO in the dispersive part of the amplitude   
means one-loop and ${\cal O} (p^4)$  
in CHPT while  NLO  in the absorptive part of the amplitude  
means two-loops and ${\cal O} (p^6)$ in CHPT. 

  In  \cite{GPS03} we took the $\pi\pi$   
scattering results from  
\cite{BKM91} and calculated  the amplitudes $K\to3\pi$  
at NLO in the isospin limit. We agreed with the NLO   
$K\to3\pi$ results recently published  in \cite{BDP02}.  
 These were previously calculated in \cite{KMW90}   
and used in \cite{KDHMW92},    
but unfortunately  the analytical full results were not available there.  
 Now, there is also available the full one-loop $K\to 3\pi$ amplitudes  
including isospin breaking effects from quark masses and electromagnetic  
interactions \cite{BB04,BB05}.

In the meantime, we discovered some small errata in the published   
formulas in \cite{GPS03} which were corrected  in \cite{SGP04}.  
We correct here another  erratum in Eqs. (6.6), (6.7) and (6.9)  
 in \cite{GPS03}  
where we wrote  $\mathbb{R}$ instead of $\widetilde{\mathbb{R}}$.  
See Section \ref{FSI} in the present work for the correction.  
In addition, matrix elements  21 and 22 
in Eqs. (6.7) and (6.9) in \cite{GPS03} were  interchanged. 
None   of these errata affects the results for the   
CP-violating asymmetries nor the conclusions given in   
\cite{GPS03,SGP04,PGS05}.  
  
 The details of the use of the optical theorem to calculate the  
two-pion cut contributions  from   
 topology   A  in Figure \ref{topologies} were given in   
Appendix E of the first reference in \cite{GPS03}
--solid circle and solid square vertices include both tree and 
one-loop level. We don't repeat these details  now. 

Notice that diagrams B and C are included
in  topology A  when one takes one of the solid vertices
at tree level and the other one at one-loop.  
 We plot explicitly diagrams B and C since  we will refer
to them later.

In the first part of this work, we complete the calculation of 
all $K \to 3 \pi$ FSI phases 
at NLO in CHPT by evaluating the three-pion cuts  
in topologies C, D and E in Figure \ref{topologies} 
for $K^+\to 3\pi$. We also  provide  with 
the neutral kaon   $K_{L,S} \to 3\pi$ 
final state interactions at ${\cal O} (p^6)$
including both two- and three-pion cut contributions.
Thus, we give the  full $K\to3\pi$ FSI  phases at NLO in CHPT  
in the isospin limit for all $K\to 3\pi$ decays.   
   
 The study of  FSI in $K\to 3\pi$ at NLO  
 also became of relevance after the proposal by Cabibbo  
\cite{CAB04} to measure  the combination $a_0-a_2$ of $\pi\pi$  
scattering lengths using  the cusp effect in  the   
$\pi^0\pi^0$ spectrum at threshold in   
$K^+ \to \pi^0\pi^0\pi^+$ and $K_L\to\pi^0\pi^0\pi^0$
decay rates\footnote{The cusp effect in SU(2) $\pi\pi$
scattering was discussed in \cite{MMS97}.}.  Within this proposal,   
it has been recently presented in \cite{CI05} the effects of FSI at NLO   
using formulas dictated by unitarity and analyticity   
approximated at second order in the $\pi\pi$ scattering lengths,   
$a_i \sim 0.2$. The error was therefore  
 canonically assumed to be of order of $a_i^2$, i.e., $5\%$.  
 There, they used a second order polynomial in the relevant final
two-pion invariant  energy $s_3$  fitted to data
to describe the 
$K\to 3\pi$ vertex that enters in the formulas of the cusp effect. 
It is of interest to check  this canonical error and provide  
a  complementary analysis  of the theoretical uncertainty.

In this work, we use our NLO in CHPT results for the real part 
of $K\to 3\pi$ fitted to data to describe  the
$K\to 3\pi$ vertex that enters in formulas of the cusp effect.  
Notice that we do {\em not} want to predict
the real part of  $K\to 3\pi$ in CHPT at any order 
but to have the best possible description fitted to data.
We treat  $\pi\pi$ scattering near threshold as in   
Cabibbo's original proposal. The advantage of using CHPT formulas
for the fit to data of the real part of $K\to 3\pi$  is that 
it contains the correct singularity structure at NLO in CHPT
which can be systematically  improved by going at higher orders.  
On the other hand, this proposal is not (and cannot be) a CHPT calculation.
    
Contributions from  next-to-next-to-leading order (NNLO) SU(3) CHPT 
in the isospin limit 
are expected typically to be around $(3\sim5)\%$, so that our  
NLO results are   just a first step in  order to reduce   
the theoretical error on the determination of the combination   
$a_0-a_2$ to  the few per cent level.  
At NNLO,  one can   follow a procedure analogous to the one  we use  here  
to get a more accurate  
measurement of $a_0-a_2$ and check the assumed NNLO uncertainty.  
At that point, and in order to reach the few per cent level in the  
theoretical uncertainty,   
it will be necessary   to include full isospin breaking effects 
at NLO too. These are also expected to be  
of a few per cent as was found for $K\to 3\pi$ in  \cite{BB04,BB05}.  
  
In Sections \ref{nota} and \ref{FSI} we introduce the needed  
notation in the description of $K\to 3\pi$ in CHPT and the  
$K^0 \to 3 \pi$ final state interaction phases  at NLO, respectively.
This completes  the $K^+ \to 3 \pi$ FSI phases already 
presented in \cite{GPS03}.  
In Section \ref{Cabibbo},  we give the relevant formulas 
of the $K \to 3\pi$  cusp effect using our NLO
formula  for the real part of $K \to 3\pi$ fitted to data.  
These could be used  in a fit of the cusp effect
to data in order to extract the $a_0-a_2$ scattering lengths combination
as pointed out in \cite{CAB04}.  We compare our results  
with the ones in \cite{CI05} and give estimates of the theoretical
uncertainties in the determination of $a_0-a_2$.
We do this both for $K^+\to \pi^0\pi^0\pi^+$ and for
$K_L\to \pi^0\pi^0\pi^0$. We would like to advance here
that our analytical results fully agree  with those in \cite{CI05}
if we make the same approximations.
Finally, we summarize our results and main conclusions in Section   
\ref{SC}.  

In Appendix \ref{inputs}  we list the numerical inputs used.  
In Appendix \ref{twopion} we present 
the neutral kaon $K \to 3 \pi$ decay FSI results from 
two-pion cut contributions.
Finally, in Appendix \ref{threepion} we collect the results for 
all three-pion cut contributions.  

\section{Notation}  
\label{nota}   
   
Here, we want to give the basic notation that we use in the following.   
In \cite{GPS03} we calculated the amplitudes  
 with $K_{1(2)} \equiv \left( K^0 -(+) \overline{K^0}\right)/\sqrt 2$,  
\ba  
\label{defdecays}  
K_2(k)&\to&\pi^0(p_1)\pi^0(p_2)\pi^0(p_3)\,,\quad [A^2_{000}]\,,
\nonumber\\  
K_2(k)&\to&\pi^+(p_1)\pi^-(p_2)\pi^0(p_3)\,,\quad [A^2_{+-0}]\,,
\nonumber\\  
K_1(k)&\to&\pi^+(p_1)\pi^-(p_2)\pi^0(p_3)\,,\quad [A^1_{+-0}]\,,
\nonumber\\  
K^+(k)&\to&\pi^0(p_1)\pi^0(p_2)\pi^+(p_3)\,,\quad [A_{00+}]\,,\nonumber\\  
K^+(k)&\to&\pi^+(p_1)\pi^+(p_2)\pi^-(p_3)\,,\quad [A_{++-}]\,,  
\ea  
as well as their CP-conjugated decays   
at NLO (i.e. order $p^4$ in this case)   
in  the  chiral expansion and in the isospin symmetry limit $m_u=m_d$. 
Disregarding CP-violating effects, that are negligible in the results 
reported in this work, $K_1\simeq K_S$ and $K_2\simeq K_L$. 
  
The lowest order SU(3) $\times$ SU(3)  
 chiral Lagrangian describing $|\Delta S|=1$ transitions is  
\ba  
\label{deltaS1}  
{\cal L}^{(2)}_{|\Delta S|=1}&=&  
C\,F_0^6 \, e^2 \, G_E \, \tr \left( \Delta_{32} u^\dagger Q u\right) 
+ C F_0^4 \left[ G_8 \, \tr \left( \Delta_{32} u_\mu u^\mu \right)  
+ G_8' \tr \left( \Delta_{32} \chi_+ \right) \right.  
\nonumber \\  &+& \left.   
G_{27} \, t^{ij,kl} \, \tr \left( \Delta_{ij} u_\mu \right) \,  
\tr \left(\Delta_{kl} u^\mu\right) \right] + {\rm h.c.}  
\ea  
with  
\be \label{Cdefinicion}  
C= -\frac{3}{5} \frac{G_F}{\sqrt 2} V_{ud} {V^*_{us}}   
\simeq -1.07 \times  10^{-6} \, {\rm GeV}^{-2}\,.  
\ee  
The correspondence with the couplings $c_2$ and $c_3$ of   
\cite{KMW90} is   
\ba  
c_2=C F_0^4 \, G_8; \nonumber \\  
 c_3= -\frac{\dis 1}{\dis 6} C F_0^4 \, G_{27} \, .   
\ea  
$F_0$ is the chiral limit value of the pion decay  
constant $f_\pi= (92.4 \pm 0.4)$ MeV,   
\ba  
u_\mu \equiv i u^\dagger (D_\mu U) u^\dagger = u_\mu^\dagger \; ,   
\nonumber \\  
\Delta_{ij}= u \lambda_{ij} u^\dagger\; (\lambda_{ij})_{ab}\equiv  
\delta_{ia} \delta_{jb}\; , \nonumber \\   
\chi_{+(-)}= u^\dagger \chi u^\dagger +(-)u \chi^\dagger u    
\ea   
$\chi= \mbox{diag}(m_u,m_d,m_s)$ is
a 3 $\times$ 3 matrix collecting the light quark masses,   
$U\equiv u^2=\exp{(i\sqrt 2 \Phi /F_0)}$ is the exponential  
representation incorporating the octet of light pseudo-scalar mesons  
in the SU(3) matrix $\Phi$;   
  
\ba  
\Phi\equiv \left(    
\begin{array}{ccc}  
\frac{\dis \pi^0}{\dis \sqrt{2}} +   
\frac{\dis \eta_8}{\dis \sqrt{6}} & \pi^+ & K^+   
\nonumber \\   
\pi^- & -\frac{\dis \pi^0}{\dis \sqrt{2}}   
+ \frac{\dis \eta_8}{\dis \sqrt{6}} & K^0   
\nonumber \\   
K^- & \overline{K^0} &- 2 \frac{\dis \eta_8}{\dis \sqrt{6}}    
\end{array}  
\right) \, .  
\ea  
  
The non-zero components of the   
SU(3) $\times$ SU(3) tensor $t^{ij,kl}$  are  
\ba  
t^{21,13}=t^{13,21}=\frac{1}{3} \, ; \, &  t^{22,23}=t^{23,22}=  
-\frac{\dis 1}{\dis 6} \, ;  
\nonumber \\  
t^{23,33}=t^{33,23}=-\frac{1}{6} \, ; \,   
&  t^{23,11}=t^{11,23}=\frac{\dis 1}{\dis 3} \, ;   
\ea  
and $Q=\mbox{diag}(2/3,-1/3,-1/3)$ is a 3 $\times$ 3  
matrix which collects the electric charge of the three light   
quark flavors.

Making use of the Dalitz variables   
\ba \label{Dalitzvar}  
x\equiv \frac{s_1-s_2}{m_{\pi^+}^2}& \hspace*{0.5cm} {\rm and}  
\hspace*{0.5cm} &  
y \equiv \frac{s_3-s_0}{m_{\pi^+}^2} \,   
\ea  
with   
$s_i\equiv(k-p_i)^2$, $3s_0\equiv m_K^2 + m_{\pi^{(1)}}^2+   
m_{\pi^{(2)}}^2+m_{\pi^{(3)}}^2$,  
the amplitudes in (\ref{defdecays}) [without isospin breaking terms]  
can be written as expansions in powers of $x$ and $y$  
\ba \label{amp1}  
A_{++-} &=& (-2\alpha_1+\alpha_3)\,-\,(\beta_1-\frac{1}{2}\beta_3+\sqrt 3  
\gamma_3)\,y + \order(y^2,x)\, ,\nonumber\\  
A_{00+} &=& \frac{1}{2}(-2\alpha_1+\alpha_3)\,-\,(-\beta_1  
+\frac{1}{2}\beta_3+\sqrt 3\gamma_3)\,y 
+ \order(y^2,x)\,,\nonumber\\  
A^2_{+-0} &=& (\alpha_1+\alpha_3)^R\,- \,(\beta_1+\beta_3)^R\,y  
 \,+\,\order(y^2,x)\, ,\nonumber\\  
A^1_{+-0} &=&  (\alpha_1+\alpha_3)^I\,- \,(\beta_1+\beta_3)^I\,y  
 \,+\,\order(y^2,x)\, ,\nonumber\\  
A^2_{000} &=& 3\,(\alpha_1+\alpha_3)^R \,+\,\order(y^2,x) \, ,\nonumber\\  
A^1_{000} &=& 3\,(\alpha_1+\alpha_3)^I \,+\,\order(y^2,x) \, ,  
\ea  
where the parameters $\alpha_i$, $\beta_i$ and $\gamma_i$ are    
functions of the  pion and kaon masses, $F_0$,   
the lowest order $\Delta S=1$ Lagrangian  
couplings $G_8$,  $G_8'$, $G_{27}$, $G_E$ and the counterterms  
appearing   
at order $p^4$, i.e., $L_i's$ and $\widetilde K_i's$.   
The definition of  these last ones  
can  be found in Section 3.2 of \cite{GPS03}.  
In (\ref{amp1}), super-indices $R$ and $I$ mean that either the real   
or the imaginary part  of the counterterms appear.

If we do not consider FSI, the complex parameters   
$\alpha_{i}$,  $\beta_{i}$  and $\gamma_{i}$  can be written at NLO   
in terms of the order $p^2 $ and $p^4$   
counterterms and the constants  $B_{i,0(1)}=  
B_{i,0(1)}^{(2)}+B_{i,0(1)}^{(4)}$   
and $H^{(4)}_{i,0(1)}$  defined in Appendix B  
of \cite{GPS03}. There, we gave $\alpha_i$, $\beta_i$ and  
$\gamma_i$ at LO in CHPT as well as  their analytic expressions  
at NLO.  
  
\section{FSI Phases for $K\to 3\ \pi$ at NLO in CHPT}  
\label{FSI}  
  
The strong FSI  mix the two final states with isospin $I=1$  
and leave unmixed  the isospin   
$I=2$ state. The mixing in the isospin $I=1$   
decay amplitudes is taken into account by introducing the  
strong re-scattering 2 $\times$ 2 matrix $\mathbb{R}$.  
The amplitudes in (\ref{defdecays}) including  the FSI effects   
can be written at all orders, in the isospin symmetry  limit,  
as follows \cite{DIPP94},    
\ba  
\label{Rescat}  
T_c\left(\begin{array}{c}A_{++-}^{(I=1)}  
\\A_{00+}^{(I=1)}\end{array}\right)_{\rm Res} &=&   
\Big(\, {\mathbb{I}}   
+i\,\mathbb{R}\,\Big) T_c\left(\begin{array}{c}A_{++-}^{(I=1)}  
\\A_{00+}^{(I=1)} \end{array}\right)_{\rm NRes} \, ,\nonumber\\  
T_n\left(\begin{array}{c}A_{+-0}^{2}  
\\ A_{000}^{2} \end{array}\right)_{\rm Res} &=&   
\Big(\,{\mathbb{I}}+i\,\mathbb{R}  
\,\Big)T_n\left(\begin{array}{c}A_{+-0}^{2}  
\\ A_{000}^{2}\end{array}\right)_{\rm NRes} \, ,\nonumber\\  
A_{++-}^{(I=2)}|_{\rm Res} &=&  \left(\,1\,+\,i\,\delta_2\,\right)
A_{++-}^{(I=2)}  
|_{\rm NRes}\, ,\nonumber\\  
\ea  
with the matrices  
\ba  
T_c\,=\,\frac{1}{3}\,\left(\begin{array}{cc}1&1\\1&-2 \end{array}\right)  
\,,\hspace{2 cm}   
T_n\,=\,\frac{1}{3}\,\left(\begin{array}{cc}0&1\\-3&1 \end{array}\right)  
\ea  
projecting the final state with $I=1$ into the symmetric--non-symmetric   
basis \cite{DIPP94}. The subscript Res (NRes) in 
(\ref{Rescat}) means that the re-scattering  
effects  have (not) been included.   
In these definitions, the matrix $\mathbb{R}$, $\delta_2$   
and the amplitudes $A^{(i)}$ depend on $s_1$, $s_2$ and $s_3$.   
  
Using the usual isospin decomposition  of $K\rightarrow 3 \pi$   
amplitudes in the isospin limit  
\ba  
A_{++-}(s_1,s_2,s_3)&=& 2 A_c(s_1,s_2,s_3) +  B_c(s_1,s_2,s_3)
+ B_t(s_1,s_2,s_3) \ , \nonumber \\  
A_{00+}(s_1,s_2,s_3)&=&  A_c(s_1,s_2,s_3) -  B_c(s_1,s_2,s_3) 
+ B_t(s_1,s_2,s_3) \ , \nonumber \\  
A_{+-0}(s_1,s_2,s_3)&=&  C_0(s_1,s_2,s_3) + 
\fr{2}{3}\Big\lbrack  B_t(s_3,s_2,s_1) -B_t(s_3,s_1,s_2) \Big\rbrack  
\nonumber \\  
&+&  A_n(s_1,s_2,s_3)-B_n(s_1,s_2,s_3) \ , \nonumber \\  
A_{000}(s_1,s_2,s_3)&=& 3  A_n(s_1,s_2,s_3) \ ,  
\ea  
 one finds for the elements of $\mathbb{R}$,  
\ba  
\mathbb{R}_{11} &=& \fr{ B_n^{\rm NRes}\,   
{\rm Im} \, A_c-B_c^{\rm NRes} \, {\rm Im}  
 \,  A_n}{A_c^{\rm NRes}B_n^{\rm NRes}-A_n^{\rm NRes}B_c^{\rm NRes}}\ ,  
\nonumber\\  
\mathbb{R}_{12} &=& \fr{ A_c^{\rm NRes}  
\, {\rm Im} \, A_n-A_n^{\rm NRes} \, {\rm Im}  
 \,  A_c}{A_c^{\rm NRes}B_n^{\rm NRes}-A_n^{\rm NRes}B_c^{\rm NRes}}\ ,  
\nonumber\\  
\mathbb{R}_{21} &=& \fr{ B_n^{\rm NRes}  
\, {\rm Im} \, B_c-B_c^{\rm NRes} \, {\rm Im}  
  \, B_n}{A_c^{\rm NRes}B_n^{\rm NRes}-A_n^{\rm NRes}B_c^{\rm NRes}}\ ,  
\nonumber\\  
\mathbb{R}_{22} &=& \fr{ A_c^{\rm NRes}  
\, {\rm Im} \, B_n-A_n^{\rm NRes} \, {\rm Im}  
  \, B_c}{A_c^{\rm NRes}B_n^{\rm NRes}-A_n^{\rm NRes}B_c^{\rm NRes}}\ .  
\ea  
At LO, $\mathbb{R}_{12}=0$ due to the fact   
that the only contributions  to   
${\rm Im} A_i$ come from  $\pi\pi$ re-scattering. At higher   
orders there are other origins for the
  contributions to  ${\rm Im} \, A_i$,
therefore $\mathbb{R}_{12}\neq 0$.  
Notice that the re-scattering matrix  $\mathbb{R}$   depends on energy 
 and at  $\pi\pi$ threshold it changes --this is in fact the cusp effect  
discussed in  the following sections.   
  
As explained in \cite{GPS03}, we used the optical theorem   
to calculate the two-pion cut  contributions to FSI phases at NLO  
for  the charged kaon decays  which are the dominant  ones.  
The analytical result of these contributions  
 for the dispersive part of the amplitudes   
for $K^+ \to 3\pi$ can be found in Appendix E  of the first reference  
in \cite{GPS03}.   
Here, we also calculate analytically these contributions to  
 FSI at NLO for  all $K_{L,S} \to 3\pi$  and  include   
the three-pion cut contributions in topologies  C, D,  and E in Figure  
 \ref{topologies}  in all cases.   
 These last three-pion cut contributions have been evaluated 
 numerically --see Appendix \ref{threepion}.  
  
The calculation has been done in the isospin limit,    
apart of the kinematical factors in the optical theorem
which are taken physical. This constitutes the   
first full calculation of FSI phases in $K\to3\pi$ decays at NLO in CHPT.  
The results for the dispersive  part of the amplitudes  
for $K_{L,S} \to 3\pi$ can be found in Appendix \ref{twopion}  
for two-pion cuts and in  Appendix \ref{threepion} for  
three-pion cuts for both neutral and charged $K\to 3\pi$.  
 For  numerical applications in the rest of  
the paper, we use the inputs in Appendix \ref{inputs}.  
  
Equations (\ref{Rescat}) imply the following relation for  
the two first coefficients of the expansion in powers of $x$ and $y$ of  
the amplitudes in (\ref{amp1})   
\ba \label{tildeRdef}  
\left(\begin{array}{c}-\alpha_1+\frac{1}{2}\alpha_3  
\\-\beta_1+\frac{1}{2}\beta_3 \end{array}\right)_{\rm Res} &=&   
\left(\,{\mathbb{I}}+ i\,\widetilde{\mathbb{R}}\,\right)  
\left(\begin{array}{c}-\alpha_1+\frac{1}{2}\alpha_3  
\\-\beta_1+\frac{1}{2}\beta_3 \end{array}\right)_{\rm NRes}\,,  
 \nonumber\\  
\left(\begin{array}{c}\alpha_1+\alpha_3  
\\ \beta_1+\beta_3 \end{array}\right)_{\rm Res} &=&   
\left(\,{\mathbb{I}}+i\,\widetilde{\mathbb{R}}\,\right)  
\left(\begin{array}{c}\alpha_1+\alpha_3  
\\ \beta_1+\beta_3 \end{array}\right)_{\rm NRes} \,,\nonumber\\  
\gamma_{3,{\rm Res}} &=&    
\left(\,1\,+\,i\,\delta_2\,\right)\gamma_{3,{\rm NRes}}\,.  
\nonumber\\  
\ea  
The matrix $\widetilde{\mathbb{R}}$ and the phase  
$\delta_2$  were given at LO in   
\cite{GPS03}.  We also gave there the  
two combinations of the  $\widetilde{\mathbb{R}}$   
matrix elements that can be obtained from the charged kaon decays.   
  As we already said in the Introduction,   
there is an erratum in Eqs. (6.6), (6.7) and (6.9) in \cite{GPS03}  
where we wrote  $\mathbb{R}$ instead of $\widetilde{\mathbb{R}}$.  
In addition, in Eqs. (6.7) and (6.9) the matrix elements  
21 and 22 were  interchanged.   
  
\section{Cabibbo's Proposal at NLO}  
\label{Cabibbo}  
  
 Recently, Cabibbo showed \cite{CAB04}  that the   
cusp effect in  the total energy spectrum of the $\pi^0\pi^0$  
pair  in $K^+ \to \pi^0\pi^0\pi^+$   
is proportional to the scattering lengths combination  
$a_0-a_2$ and proposed to use this effect to measure it.    
  
 This interesting  proposal was done at lowest order   
in the sense that the author just considered topologies of type  
A in Figure \ref{topologies} with one $\pi\pi$ scattering vertex
in the final state. It has been  followed up  
by a study of higher order re-scattering effects coming from  
topologies B and C in Figure \ref{topologies} \cite{CI05}.  
 The first experimental analysis   
applying this proposal has been already  published in \cite{NA48} 
showing   clearly that the  nominal 5 \% theoretical accuracy  quoted in   
\cite{CI05}  will dominate this determination. 
It is therefore very important to check   
this 5\% theoretical uncertainty and how to reduce it further.  
  
 In \cite{CI05}, in order to make a quantitative  evaluation for the   
 Cabibbo's proposal uncertainties, a power  counting  in  
the scattering lengths $a_i$ was done. The main   
conclusion was that one needs to include topology C  
in Figure \ref{topologies}   to go to ${\cal{O}}(a^2)$ accuracy,   
i.e. around $5\%$.  
   The authors used Feynman diagrammatics to do the
counting in powers of $a_i$  although they  did not have  an effective
 field theory supporting it.  
It is of course interesting to construct it and
follow that program.  In fact, very recently,
 such an effective field theory has been presented \cite{GAS06}. 
  
In \cite{CAB04,CI05}, the  
real part of the $K \to 3 \pi$ amplitude
was approximated by a second order polynomial in the relevant final
two-pion invariant  energy $s_3$  fitted to data.
We  study a variation of 
Cabibbo's proposal  that uses NLO CHPT formulas
for the real part of $K\to 3\pi$ vertex fitted to data instead of the
 polynomial approximation used in \cite{CAB04,CI05}
plus analyticity and unitarity. 

 At a given order in the $\pi\pi$ re-scattering,
 any FSI diagram with at least one two-pion cut can be 
drawn as  topology A in  Figure \ref{topologies}.   
Solid square vertex stands for  the effective $K \to 3\pi$ decay
vertex and  solid circle vertex stands for the effective 
$\pi\pi$ scattering. 
FSI diagrams with no two-pion cuts  are treated apart.
At NLO, these are diagrams D and E in Figure \ref{topologies}. 
Using analyticity and unitarity, 
one can  cut topology A into two subdiagrams: the right-hand  side
is $\pi\pi$ scattering and the left-hand side is $K \to 3\pi$. 
Diagrams B and C are the only two that
appear  at NLO in the $\pi\pi$ final state re-scattering, i.e.
with  two  $\pi\pi$ scattering vertices and 
with at least one two-pion cut.

The cusp effect we are interested in originates in the different 
contribution of $\pi^+\pi^- \to \pi^0 \pi^0$ scattering to 
the $K^+\to \pi^0\pi^0\pi^+$  (or $K_L\to \pi^0\pi^0\pi^0$)  
amplitude when the invariant $\pi^0\pi^0$
energy is above or  below $\pi^+\pi^-$ production threshold. 
 In Section  \ref{charged} for 
$K^+ \to \pi^0 \pi^0 \pi^+$  and in Section \ref{neutral} for
$K^0 \to \pi^0 \pi^0 \pi^0$, we obtain these contributions
using  just analyticity and unitarity near threshold, i.e.  applying 
 the optical theorem  to calculate the imaginary part 
of  the discontinuity across the physical $\pi^+\pi^-$ cut and 
 Cutkosky rules to calculate the real part of  that  discontinuity
around threshold.  
In some cases, one can get pieces of the real part of that
discontinuity by applying the optical theorem  
below $\pi^+\pi^-$ threshold.  Sometimes this becomes unphysical
because the value of the real world pion masses forbids it. 
In such cases we follow the strategy in \cite{CI05} of going first
 to unphysical values of pion masses, e.g.  
 $m_{\pi^0} > m_{\pi^+}$,  such that the absorptive part below threshold 
exists and  apply the optical theorem in this set up.  
This result is analytically continued above threshold where
the amplitude always exists  by putting the real value for  pion masses. 
  
 As said above, analyticity and unitarity  allows to separate   
 $\pi\pi$ scattering, i.e the   
scattering length effects, from the rest in $K \to 3\pi$.
In fact,  when  this $\pi\pi$  scattering is  evaluated  around 
 threshold,  one is  intuitively lead to use the scattering length
to all orders as a good approximation  
for the real part of  $\pi\pi$ scattering.
 Explicitly,  we follow   \cite{CAB04,CI05}  
for the treatment of $\pi\pi$ scattering matrix elements near threshold.  
 Near threshold,  we use \cite{GL84}  
\ba  
\label{scat}  
A_{00}: \hspace*{0.5cm}&  
\pi^0 \pi^0 \to \pi^0 \pi^0,   
\hspace*{0.5cm}&  
{\rm Re} \, A_{00} \equiv {32\pi \, a_{00}(s)}   \, 
\nonumber \\  
A_{+0}: \hspace*{0.5cm}&  
\pi^+ \pi^0 \to \pi^+ \pi^0,   
\hspace*{0.5cm} &  
{\rm Re} \, A_{+0} \equiv {32\pi \, a_{+0}(s)}  \, 
\nonumber \\  
A_{x}: \hspace*{0.5cm}&  
\pi^+ \pi^- \to \pi^0 \pi^0,   
\hspace*{0.5cm} &   
{\rm Re} \, A_{x\;} \equiv {32\pi \, a_{x}(s)} \, 
\nonumber \\  
A_{+-}: \hspace*{0.5cm}&  
\pi^+ \pi^- \to \pi^+ \pi^-,   
\hspace*{0.5cm} &  
{\rm Re} \, A_{+-} \equiv { 32\pi \, a_{+-}(s)}  \, 
\nonumber \\  
A_{++}: \hspace*{0.5cm}&  
\pi^+ \pi^+ \to \pi^+ \pi^+,   
\hspace*{0.5cm} &  
{\rm Re} \, A_{++} \equiv {32\pi \, a_{++}(s)}  \, , \nonumber \\ 
\ea  
as a definition
 of the different effective scattering length combinations $a_{ij}(s)$,  
which are the unknown quantities  to be obtained by fitting 
the cusp effect.   From these fitted effective scattering lengths
combinations,  
one can  extract the pion scattering lengths by comparing them
with their CHPT corresponding prediction \cite{CI05}.
Notice that this procedure is  sensible  since the effective scattering 
lengths are evaluated near threshold and chiral corrections are tractable 
within CHPT. These  include  radiative and isospin breaking corrections 
to the isospin limit results. These isospin limit results are: 
$a_{00}= (a_0+a_2)/3$, $a_{+0}=a_2/3$,
$a_x=(a_0-a_2)/3$, $a_{+-}= (2 a_0 + a_2)/3$ and
$a_{++}=a_2$.  
We {\em define} these isospin limit scattering lengths 
at the following   
thresholds $s_{th}$: $4 m_{\pi^+}^2$ for $a_{00}$, $a_x$,  
$a_{+-}$ and $a_{++}$ and $(m_{\pi^+}+m_{\pi^0})^2$ for $a_{+0}$.  
  
 In order  to have a more accurate  description of 
 $\pi\pi$ scattering near threshold
 than  the one given by (\ref{scat}),
one can follow \cite{CI05} and  
perform an expansion around threshold
in the different kinematical variables    
on which the amplitudes depend.  
 Up to linear terms, the generic matrix elements   
--neglecting all higher order but the P wave-- are of the  form   
\cite{CGL00}  
\be\label{scat2}  
{\rm Re} \, A_{ij}=32\pi\left[ a_{ij}(s)+  
\frac{3}{4} \,  a_{ij}^P  
\, \frac{ (t-u)}{m_{\pi^+}^2}\right]  
\ee  
with\footnote{The  effects   of $\pi\pi$ threshold singularities  
are NNLO in a chiral counting and expected to be small.  
 However one should include them   
since as we said we want to treat the $\pi\pi$ scattering   
non-perturbatively 
-- this  can be done as in  \cite{CI05}.}   
\be  
a_{ij}(s) = a_{ij}\left[1+r_{ij}\frac{(s-s_{th})}{4m_{\pi^+}^2}  
\right]\,  
\label{eq:aap}  
\ee  
where $a_{ij}(s)$ are the ones in  (\ref{scat}).  
For numerical applications, we use   
$r_0=1.25\pm0.04$ and $r_2=1.81\pm0.05$ \cite{CGL00}   
which are compatible with the results in \cite{PY04} and   
\ba  
a_{+-}^P= a_{+0}^P = a_1/2  
\ea  
with  $a_1^{CHPT}=m_{\pi}^2/(12\pi F_0^2)$  to lowest order in CHPT.   
The rest of $a_{ij}^P$ are zero. 
    
 The above discussion fixes  the real part of $\pi\pi$ amplitudes  
near threshold which are the unknowns
to be fitted. However, even if $s_3$ is always around threshold,
at NLO $\pi\pi$ amplitudes are also needed
far from threshold and the effective scattering lengths in (\ref{scat})  
do not give a good description of them.  As we will see in the next
sections, the two cases appear clearly separated 
in the $\pi\pi$ scattering amplitudes at  NLO 
in (\ref{ABscatNLO}) and (\ref{AB0NLO})  which
are either evaluated at $s_3$,  at  $(m_K^2+ 3 m_\pi^2 -s_3)/2$ or 
at  $(m_K^2+ 7 m_\pi^2 + s_3)/4$. 
We only leave as unknown the scattering lengths
evaluated at $s_3$ around threshold and  use 
the real part of the full $\pi\pi$ amplitudes  at NLO in CHPT 
expressions in the other cases since the scattering lengths
 are evaluated far from threshold.
The accuracy in the description of the $\pi\pi$ amplitudes 
in these  latter cases can be improved systematically by going at higher
order in CHPT.

For the other ingredient needed by unitarity, 
i.e. the real part of  $K\to 3\pi$ amplitudes,
  the procedure  we propose, in order   
to take into account the singularity structure in
the real part of $K \to 3 \pi$ amplitudes in a systematically   
improvable manner, is   to treat this real part of $K \to 3 \pi$ 
  within CHPT.    I.e., first using the  tree-level (LO)
CHPT formulas,    next the  one-loop (NLO)ones,  $\cdots$;
always fitted to data.  This is a
difference with the approach in \cite{CI05} where a second
order polynomial is used. We want to quantify the numerical
differences induced by this 
in  the scattering lengths obtained from the cusp.
There are also other differences coming from  some approximations 
done in \cite{CI05}. These differences and their effects
will become clear in the  following sections.

 Notice that the use of the CHPT formula
for the real part of $K\to 3\pi$ does  
not change the need of fitting it to data.   
The aim is {\em not} to predict the real part of $K\to 3 \pi$ but   
to describe its analytic structure as accurately  
 as possible so that, once separated the real part of
$K\to 3\pi$  from  $\pi\pi$ scattering through the optical theorem
 and Cutkosky rules, one can measure $\pi\pi$ scattering near threshold 
as Cabibbo proposed.  In particular, $\pi\pi$ scattering 
near threshold is treated non-perturbatively as in \cite{CAB04}.
The advantage  now is that we have
the correct structure of singularities  at a given CHPT order for
the real part of $K\to 3\pi$.  
Below we give the formulas which describe the 
cusp effect at NLO within this approach.  
  
\subsection{Cabibbo's Proposal for Charged Kaon Decays}  
\label{charged}  
  Near $\pi^+\pi^-$ threshold, 
we can decompose the $K^+\to\pi^0\pi^0\pi^+$ amplitude 
as follows \cite{CAB04,CI05}  
\ba\label{ABdefinition}  
A_{00+} &=&\left\{
\begin{array}{ll}
 \overline A_{00+} +   
\overline B_{00+}v_{\pm}(s_3), &\quad {\rm for} \quad  s_3 > 4m_{\pi^+}^2  
\\
 \overline A_{00+} +   
i\overline B_{00+}v_{\pm}(s_3) , &\quad {\rm for} 
\quad s_3 < 4m_{\pi^+}^2  ,
\end{array} \right. \nonumber \\
\ea  
where $\overline A_{00+}$ and $\overline B_{00+}$ are 
in general singular functions except near the $\pi^+\pi^-$ threshold 
\cite{CGKR05} and 
\be  
v_{ij}(s)=\sqrt{\frac{\vert s-(m_{\pi^{(i)}}+m_{\pi^{(j)}})^2\vert}{s}}.  
\ee  
Notice that these kinematical factors are taken with  physical pion
masses, in this way one can describe the cusp effect which is 
 generated by the different behavior  
for the final two neutral pions invariant energy  above and below  the  
$s_3= 4 m_{\pi^+}^2$ threshold.
 
With these definitions, the differential decay rate for this amplitude   
can be written as \cite{CI05}  
\ba  
\label{tot}  
 | A_{00+}|^{2} \equiv    
\re \overline A_{00+}^2 + \Delta_{A} + v_{\pm}(s_3)   
\Delta_{\rm cusp} \,,  
\ea
with  
\ba  
 \Delta_{A} &\equiv&
 \im \overline A_{00+}^2 + v^2_{\pm}(s_3)   
\left\lbrack  
\re \overline B_{00+}^2 + \im \overline B_{00+}^2 \right \rbrack \, ,\\  
\Delta_{\rm cusp} &\equiv & 
 \left\{   
\begin{array}{ll}    
      -2 \re \overline A_{00+}\im \overline B_{00+} +   
2\im \overline A _{00+}\re \overline B_{00+} \, , 
 \quad {\rm for} \,\,\,  s_{3}<4m_{\pi^{+}}^{2};&  \\  
      2 \re \overline A_{00+}\re \overline B_{00+} +    
2 \im \overline A_{00+}\im \overline B_{00+}\, , 
 \quad   {\rm for} \,\,\,  s_{3} > 4m_{\pi^{+}}^{2} . & \end{array} 
\right.   
\label{Dcusp}  
\ea
The combination of real and imaginary amplitudes $\Delta_{\rm cusp}$   
defined above parametrizes the cusp effect due to the   
$\pi^+\pi^-\to\pi^0\pi^0$   
re-scattering in the $K^+\to\pi^0\pi^0\pi^+$    decay rate.

   The real part of the amplitude $K^+ \to \pi^0 \pi^0 \pi^+$,   
i.e. Re $\overline A_{00+}$,  was calculated at NLO   
in CHPT in the isospin limit in \cite{GPS03,BDP02}.    
 As said before, we don't want to predict it (or any NLO counterterm)
but to use  Re $\overline A_{00+}$ at NLO in CHPT formula
fitted  to data in  $\Delta_{\rm cusp}$. 
 This provides  a fit at least as good as the second order polynomial 
parametrization used   in the original proposal \cite{CAB04,CI05},
 and its precision 
is just limited by data and the singularity structure at NLO in CHPT.

We get $ \im \overline A_{00+}$ and  $\im \overline B_{00+}$   
using the optical theorem,  where we   
consistently use the   NLO expression for 
$\rm \re \, \overline  A_{00+}$  previously fitted to data. 

The singular part of the dispersive amplitude, i.e. 
${\rm Re}\overline B_{00+}v_{\pm}(s_3)$, 
is proportional near threshold
to  one half of the discontinuity across the physical cut
\footnote{We thank J\"urg Gasser for clarifying this point to us.}
which is also fixed just by unitarity and analyticity.

 With those results at hand, we can study  the proposal
of determining the scattering lengths combination 
$a_0-a_2$  from  (\ref{Dcusp}).

 The real part of $\pi\pi$ scattering in the final state interactions
is  included as in \cite{CI05}, using just unitarity and analyticity
and its treatment was 
explained in the introduction of Section \ref{Cabibbo}.
In particular, for the real part of $\pi^+\pi^-\to\pi^0\pi^0$ scattering
near threshold, we use the non-perturbative definition  
$a_x(s_3)$ in (\ref{scat}).

 First, we give the expressions for   
$\overline A_{00+}$  and $\overline B_{00+}$
using the tree-level CHPT expression to fit the real part of 
$K^+\to \pi^+\pi^0\pi^0$ to data. We get
\ba \label{ABscatLO}  
\re{\overline A_{00+}}\vert_{LO} (s_3)  
&=& C'\Big[ \re \, G_8(\vpd-s_3)+\frac{G_{27}}{6(\vkd-\vpd)}  
\left[s_3(4\vpd-19\vkd)\right.\nonumber\\
&&\left.+5\vkc-4\vpc+19\vkd\vpd\right]\Big]\, ,\nonumber\\  
\im{\overline A_{00+}}\vert_{LO} (s_1,s_2,s_3)  
&=& v_{00}(s_3) a_{00}(s_3) \, 
\re{\overline A_{00+}}\vert_{LO} (s_3)  
\nonumber \\ &+&  \Big\lbrack v_{+0}(s_1) 
(SW_{+0}^{LO}(s_1)+(s_2-s_3) PW_{+0}^{LO}(s_1))+  
 s_1\leftrightarrow s_2  \Big\rbrack \, ,   
\nonumber \\  {\rm with} \quad \quad
SW_{+0}^{LO}(s)&=& C' \, \frac{(s-2 m_\pi^2)}{64\pi f_\pi^2} 
\Big\lbrack \re   \,  G_8(m_K^2 + m_\pi^2 -  
 s) \nonumber \\ &+&
\frac{G_{27}}{6(m_K^2-m_\pi^2)}(9 m_K^4 
 +m_K^2 (15 m_\pi^2 - 19 s)   
+ 4 m_\pi^2 (s-m_\pi^2))\Big\rbrack \nonumber \\  
{\rm and} \quad \quad
PW_{+0}^{LO}(s)&=& C' \, 
\frac{(s- 4 m_\pi^2)}{192\pi f_\pi^2}  \Big\lbrack  
\re \, G_8+\frac{G_{27}}{6(m_K^2-m_\pi^2)} 
(19 m_K^2-4 m_\pi^2)\Big\rbrack  
\,,\nonumber\\   
\re{\overline B_{00+}}\vert_{LO} (s_3)  &=& 0\, , \\  
\im{\overline B_{00+}}\vert_{LO} (s_3)  &=& C'  
a_x(s_3)\left\lbrack   
-\re \, G_8(s_3-\vpd+\vkd)+ G_{27}\frac{1}{3}\left[13s_3-7\vkd  
-13\vpd\right]\right\rbrack\, , \nonumber 
\ea  
where $C'=i \, C \, F_0^4/ (f_{\pi}^3 \, f_K)$ 
and the constant $C$ is defined in    (\ref{Cdefinicion}). 
Notice that $LO$ here does not mean a calculation at LO in CHPT. 
It means  that we have consistently used the LO CHPT expression to fit
$\re \overline A_{00+}$, while $ \im \overline A_{00+}$ and  
$\im \overline B_{00+}$  are obtained also
at LO in $\pi\pi$ scattering using the optical theorem.
 The $SW$ and $PW$ pieces above are  the S wave and P wave
contributions to $\im \overline A_{00+}$, respectively.  

At NLO in $\pi\pi$ scattering
and using the CHPT expression to one-loop \cite{GPS03,BDP02} 
to fit the real part of $K^+ \to \pi^0\pi^0\pi^+$ to data,
we get  
\ba \label{ABscatNLO}  
\re{\overline A_{00+}}\vert_{NLO} (s_1,s_2,s_3)  
&=& \re{\overline A_{00+}}\vert_{LO} \,+\,  
\Big[ M_7(s_3)+M_8(s_1)+M_8(s_2)+M_9(s_1)(s_2-s_3) 
\nonumber\\ &&  
+M_9(s_2)(s_1-s_3)\big]_{\order(p^4)}\, + 
\delta \re {\overline A_{00+}} (s_1,s_2,s_3)\, , \nonumber\\  
\im{\overline A_{00+}}\vert_{NLO} (s_1,s_2,s_3)  
&=& \im{\overline A_{00+}}\vert_{LO}  
\, +\,v_{00}(s_3)a_{00}(s_3)\left\lbrack  
  M_{7}(s_3)+\widetilde{M}_{8}(s_3)  \right.  \nonumber \\  &&  
+\widetilde{M}_{9}(s_3) (m_K^2+3 m_\pi^2-2  s_3)  
\left. -\widetilde{M}_{9}^s(s_3)\right]_{\order(p^4)}  \nonumber\\  
&&+ \Big\lbrack v_{+0}(s_1)(SW_{+0}^{NLO}(s_1)+  
(s_2-s_3) PW^{NLO}_{+0}(s_1)) +  
s_1 \leftrightarrow s_2 \Big\rbrack\,,\nonumber \\  
\re{\overline B_{00+}}\vert_{NLO} (s_3)
&= &- a_{00}(s_3)\, v_{00}(s_3)\, 
\im{\overline B_{00+}}(s_3)   \nonumber\\
&&-  a_x(s_3)^2 \, v_{00}(s_3)\, {\displaystyle \int_{-1}^1} \,
{\rm d} x  \, \re{\overline A_{00+}}
\left(t^+(s_3,x), t^-(s_3,x), \, s_3\right)\,  \nonumber\\
&& - a_x(s_3) \,
 \frac{1}{2} \, {\displaystyle\int_{-1}^1} \, {\rm d} x \,
\im  \widetilde A_{++-}  (t^+(s_3,x), t^-(s_3,x))  \, ,\nonumber \\  
\im{\overline B_{00+}}\vert_{NLO} (s_3)  
&=&\im{\overline B_{00+}}
\vert_{LO}\,+\,  a_x(s_3)\Big[ 2 M_{11}(s_3)
+ \widetilde{M}_{10}(s_3)+ \widetilde{M}_{11}(s_3)   
 \nonumber \\  && - \widetilde{M}_{12}(s_3) 
 (m_K^2+3 m_\pi^2-2  s_3) + \widetilde{M}_{12}^s(s_3)
\Big]_{\order(p^4)}  \,,  
\ea  
when  $s_3$ is near $4 m_{\pi^+}^2$.
Where $t^\pm(s,x)=a(s)\pm b(s) x$ with  
$a(s)$  and $b(s)$ defined in (\ref{def_a_b}). 
We have not included the three-pion cut graph 
contributions   which we have checked  (see Appendix 
\ref{threepion} for the results) to be  negligible.
 
The S and P wave contributions  to $\im \overline A_{00+}$ 
--$ SW_{+0}^{NLO}(s_1)$    and $PW^{NLO}_{+0}(s_1)$--   were
calculated in \cite{GPS03}. With  the substitution $\sigma(s_i)
\rightarrow v_{+0}(s_i)$  in the formulas  of \cite{GPS03} understood,
one has  
\ba \label{SWPW}  
 v_{+0}(s_1) SW_{+0}^{NLO}(s_1)+ s_1 \leftrightarrow s_2 &=& 
{\rm formulas\; E.25\;   
+\; E.29\; of\; [4] }  \ ,
\nonumber \\  
 v_{+0}(s_1)(s_2-s_3) PW_{+0}^{NLO}(s_1)+ s_1 \leftrightarrow s_2 &=&  
{\rm formulas\; E.26\;   
+\; E.30\; of\; [4] }  \ .
\ea  

  The  functions $M_i(s)$ used in (\ref{ABscatNLO})   
are the same ones defined in \cite{BDP02} at order $p^4$ 
with the exception of  $B(m_{\pi},m_{\pi},s_3)$ which has
to be exchanged  by   
\be \label{newB}  
{\cal B}(m_{\pi},m_{\pi},s_3) = {\cal J} (m_{\pi},s_3) -\frac{1}{16\pi^2}  
\left[\log{\left(\frac{m_{\pi}^2}{\nu^2}\right)}+1\right]  
\ee  
with   
\ba  
{\cal J}(m_{\pi},s_3) &=& \frac{1}{16\pi^2}\left\lbrace   
\begin{array}{ll}  
2+v(s_3)\log{\left(\frac{1-v(s_3)}  
{1+v(s_3)}\right)}, \quad {\rm for}
\quad  s_3 > 4m_{\pi^+}^2;\nonumber\\  
2+2 \,v(s_3)   
\arctan \left(v(s_3)\right), \quad {\rm for}
\quad s_3 < 4m_{\pi^+}^2;  
\end{array}  
\right.  
\ea  
and $v(s_3)=\sqrt{\vert s_3-4m_{\pi}^2\vert/s_3}$.    
Functions $\widetilde{M}_i(s)$ and $\widetilde{M}_i^s(s)$ are defined in 
(\ref{tildeMis}) as integrations on $x$ of
$M(t^+(s,x))$ and $t^+(s,x) \, M(t^+(s,x))$, respectively. 

Some pieces of $\rm Re \, \overline B_{+00}$ can be obtained
going below $\pi^+\pi^-$ threshold and applying the optical theorem there
but one cannot get all of them in such way. 
The first two lines in $\rm Re \, \overline B_{+00}$
in (\ref{ABscatNLO}) come from diagram B in Figure
\ref{topologies} while the third  line  comes from diagram
C in the same figure.

 The real part of the discontinuity  across
the physical cut in $s_3$ for $K^+ \to \pi^0\pi^0\pi^+$
can be written as an integral in the $x$-complex plane
between $x=-1$ and $x=+1$ \cite{ANI03}
\footnote{We thank J\"urg  Gasser for bringing  this work
to our attention.}.  This can be expressed
as the sum of an integral  in $x$ along the real axis
with $-1 \leq x \leq 1$  which is finite 
 plus an additional piece which is non-zero 
for values of $s_3$ above
\ba
\label{sL}
s_L &=&  \frac{m_{\pi^0}}{m_{\pi^0}+ m_{\pi^+}} \left(m_{K^+}^2
- m_{\pi^+}^2\right) \,  
\ea
 and diverges as
$s_3$ approaches  $(m_{K^+}-m_{\pi^+})^2$.
This piece takes into account the effect of the presence of the  
singularity at the pseudo-threshold
$s_3=(m_{K^+}-m_{\pi^+})^2$ \cite{ANI03} and  
gives an additional contribution 
to the result for $\rm Re \, \overline B_{+00}(s_3)$  
in (\ref{ABscatNLO}) when  $s_3$ is above $s_L$.
This additional piece is known and can be expressed
as the result of  an integral over 
a circuit  in the complex-$x$ plane around a branch cut.
Since we need  to describe the cusp   just near  threshold, this extra
piece   is not needed in $\rm Re \, \overline B_{+00}(s_3)$ 
and one can effectively use the result 
for $\rm Re \, \overline B_{+00}(s_3)$ in  (\ref{ABscatNLO})  
which is therefore finite. This agrees with the naive result of applying 
Cutkosky rules when $s_3$ below $s_L$.

We get the function  $\im  \widetilde A_{++-}(t_1,t_2)$ 
that appears in $\rm Re \, \overline B_{+00}(s_3)$ from the imaginary 
part of the $K^+\to  \pi^+ \pi^+ \pi^-$ decay amplitude,
 which was obtained in 
\cite{GPS03} using the optical theorem.  From that result  and 
disregarding the  tiny P wave contribution,
we get
\ba
\label{ImA++-}
\im  \widetilde A_{++-} (t_1,t_2) &=& 
a_x(t_1) v_{00}(t_1) {\displaystyle\int_{-1}^1} \, {\rm d} y \,
\re{\overline A_{00+}}(t^+(t_1,y) , t^-(t_1,y), t_1)
 \nonumber\\ &&
+ 2 a_{+-}(t_1) v_{\pm}(t_1)
\, {\displaystyle\int_{-1}^1} \, {\rm d} y \,
\re{\overline A_{++-}}(t^+(t_1,y), t_1, t^-(t_1,y)) \,
 \nonumber\\ &&+  \, a_{++}(t_2) v_{\pm}(t_2)
\, {\displaystyle\int_{-1}^1} \, {\rm d} y \,
\re{\overline A_{++-}}(t^+(t_2,y),  t^-(t_2,y), t_2) \, , 
\ea
where the expression for $\re{\overline A_{++-}}$ at NLO in CHPT can 
be found in \cite{GPS03,BDP02}.  This expression has to be fitted
to  $K^+ \to \pi^+\pi^+\pi^-$ data.
Notice that,  in the formula above, $\pi\pi$ amplitudes
 appear  in some cases evaluated far from threshold  
even if $s_3$ is around threshold. 
As said before, in Section \ref{Cabibbo}, these cases are 
clearly separated and 
whenever this happens we use full NLO CHPT 
predictions and not the effective scattering length combinations
in (\ref{scat}) which are used as unknowns just near threshold.

The contribution $\delta \re {\overline A_{00+}}$ 
to $\re {\overline A_{00+}}\vert_{NLO}$ in the first line
of (\ref{ABscatNLO})  
comes from the discontinuity across the physical cuts and 
takes into account the singularities  of $\re {\overline A_{00+}}$
at  $s_i=(m_{\pi^+}+m_{\pi^0})^2$  ($i=1,2$) 
and $s_3=4 m_{\pi^0}^2$ 
thresholds which start at order $p^6$ in CHPT. Using Cutkosky rules, 
we get
\ba
\label{deltaA}
\delta \re {\overline A_{00+}} (s_1,s_2,s_3) &=& 
-  \,a_{00}(s_3)v_{00}(s_3)
{\displaystyle\int_{-1}^1} \, {\rm d} x \ \, a_{+0}(t^+(s_3,x))
\, v_{+0}(t^+(s_3,x))  \nonumber \\ 
&& \times {\displaystyle \int_{-1}^1}
\, {\rm d} y \, \re \overline 
A_{00+}(t^+(s_3,x),t^+(t^+(s_3,x),y), t^-(t^+(s_3,x),y))  \nonumber\\
&&- \Big\{ \,a_{+0}(s_1)v_{+0}(s_1){\displaystyle\int_{-1}^1} \, 
{\rm d} x  
\left[ \, {\cal F}_{00+}(s_1,t^+(s_1,x),t^-(s_1,x)) 
\right. \nonumber\\
&& +v_\pm(t^-(s_1,x)) \left.  
\im \overline B_{00+}(t^-(s_1,x)) \right]
+  s_1 \rightarrow s_2 \Big\} \, \, . 
\ea
 ${\cal F}_{00+}(s_1,s_2,s_3)$ is equal to 
$\im \overline{A}_{00+}(s_1,s_2,s_3)$  in  (\ref{ABscatNLO})
minus the terms proportional to $v_{+0}(s_1)$ which produce a 
regular piece. Again, we know from the result in \cite{ANI03} that 
 (\ref{deltaA}) is just correct when  $s_3$ is in   the range 
\ba
\label{range}
4 m_{\pi^0}^2 
\leq s_{3} \leq \frac{m_{\pi^0}}{m_{\pi^0}+ m_{\pi^+}} \left(m_{K^+}^2
- m_{\pi^+}^2\right) \,  
\ea
 and both $s_1$ and $s_2$ are in the range, 
\ba
\label{range2}
(m_{\pi^0}+m_{\pi^+})^2 
\leq s_{1,2} \leq \frac{1}{2} \left(m_{K^+}^2
- m_{\pi^+}(2 m_{\pi^0}-m_{\pi^+})\right) \, . \nonumber \\
\ea
For larger values of $s_i$, the same comment around 
(\ref{sL}) applies, i.e., known additional pieces
that  diverge as $s_i$ approaches $(m_{K^+}- m_{\pi^{(i)}})^2$
 have to be added \cite{ANI03}.   
The discontinuities  in $s_1$ and $s_2$
included in  $\delta \re {\overline A_{00+}}$
are completely analogous
to the  one in $s_3$ already discussed after (\ref{ABscatNLO}).
We do not repeat therefore  the discussion
already done for the discontinuity in  $s_3$.

The difference here is that,  while  
we need the discontinuity   in $s_3$  just around its threshold, 
it is  possible to approach the pseudo-thresholds in $s_1$ 
(or in $s_2$) when $s_3$ is around threshold.
And when this  happens,  one has to take 
into account in $\delta \re {\overline A_{00+}}$
the additional  pieces mentioned above which 
 diverge at  pseudo-thresholds.
A solution  to this inconsistency  which does not simply use
the discontinuity to describe the cusp effect
is discussed in \cite{CGKR05}.
Another possible solution, if one persists in using the discontinuity
to describe the cusp effect in $s_3$, is to drop out 
$\delta \re {\overline A_{00+}}$  and use  instead
the (unknown yet) full-two loop  
finite relevant pieces  to describe the NLO 
singularities near  $s_i=(m_{\pi^+}+m_{\pi^0})^2$  ($i=1,2$) 
and $s_3=4 m_{\pi^0}^2$ thresholds. 
These additional pieces
could also be fitted to data since we don't want to predict
$\re {\overline A_{00+}}$  but to obtain the best description of it.
Here, we don't discuss these possible solutions further and leave
it for a future study. 

  In \cite{CI05},  they used the approximation that
 the integrands in the finite integrals 
(\ref{ABscatNLO}),   (\ref{ImA++-}) 
and (\ref{deltaA}) are to a good accuracy linear 
in the variables  $x$ or  $y$.
 If we make such approximation, we get
\ba
\label{reB00+}
\re \overline B_{00+}\vert_{NLO} (s_3)&=& 
-a_{00}(s_3)\, v_{00}(s_3) \, \im{\overline B_{00+}}(s_3)   \nonumber\\ &&
-   2 a_x(s_3)^2 v_{00}(s_3) \re{\overline A_{00+}}
\left(a(s_3),a(s_3),\, s_3\right)\,   \nonumber\\
&&- 2 a_x(s_3) a_x(a(s_3)) v_{00}(a(s_3))
\re{\overline A_{00+}}(a(a(s_3)), a(a(s_3)),a(s_3))
 \nonumber\\
&& - 4 a_x(s_3) a_{+-}(a(s_3)) v_{\pm}(a(s_3))
\re{\overline A_{++-}}(a(a(s_3)), a(s_3), a(a(s_3)))
\nonumber\\
&& - 2 a_x(s_3) a_{++}(a(s_3)) v_{\pm}(a(s_3))
\re{\overline A_{++-}}(a(a(s_3)), a(a(s_3)),a(s_3))\nonumber\\
\ea
and 
\ba
 \label{reA00+}
\delta \re {\overline A_{00+}} (s_1,s_2,s_3) &=& 
- 4 \,a_{00}(s_3)\,a_{+0}(a(s_3)) \, v_{00}(s_3) \,  
v_{+0}(a(s_3))\re \overline  A_{00+}(a(s_3),a(a(s_3)),a(a(s_3)))
\nonumber\\&&- 2\left\{ \,a_{+0}(s_1)v_{+0}(s_1)
\left[ {\cal F}_{00+}(s_1,a(s_1),a(s_1)) 
 +v_\pm(a(s_1)) \,  \im \overline B_{00+}(a(s_1))\right] \right.\nonumber\\
&&\left.+ s_1 \rightarrow s_2 \right\}
\ea
which agree with \cite{CI05}. By doing
these finite integrals numerically, we have checked that 
the approximation that the integrands are linear in 
 $x$ or  $y$ is good  to 1 \% accuracy.

Again, in (\ref{ABscatNLO}),
 the subscript $NLO$  does not mean a calculation at NLO in CHPT.   
It means that we have consistently used the NLO CHPT 
isospin limit expression to fit the effective
$\re \overline A_{00+}$  and included $\pi\pi$ re-scattering  
vertices also at NLO using just unitarity and analyticity.
Notice that in particular CHPT is not used to include 
$\pi\pi$ re-scattering effects.
 These are unknowns to be determined  and are treated to all orders 
near threshold.  Sometimes effective $\pi\pi$ vertices 
appear evaluated  far from threshold even if $s_3$ is around
$4 m_{\pi^+}^2$.  In these cases we use the $NLO$ CHPT 
prediction and don't leave them as unknowns.

\begin{figure}
\begin{minipage}[t]{3cm}\vskip-6.5cm 
$\small{100 \times} \frac{
{\rm d}| \Gamma_{\rm cusp}(s_3)|/{\rm d} s_3}{{\rm d} 
\Gamma(s_3)/{\rm d} s_3}$
 \end{minipage}
\includegraphics[{width=10cm}]{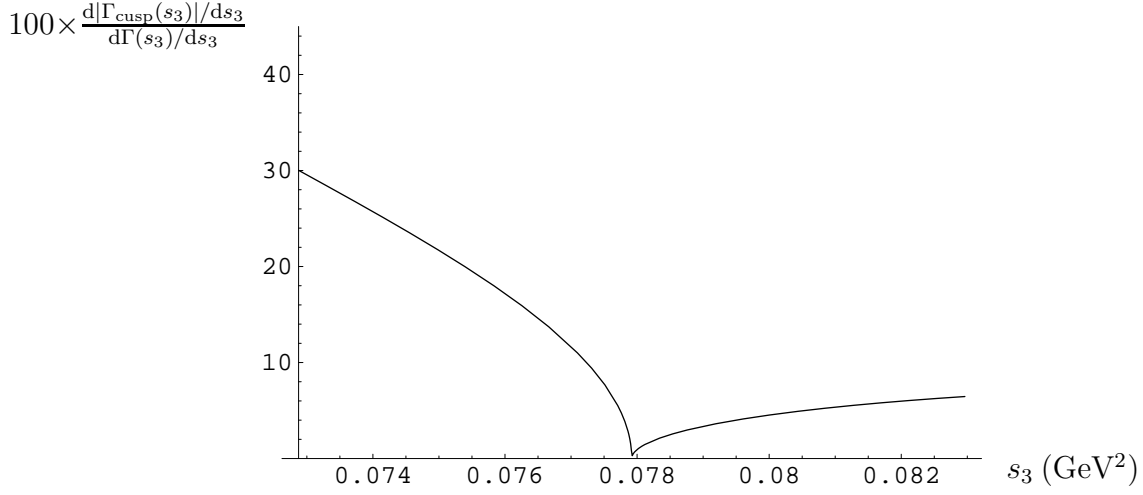}  
\begin{minipage}[t]{1cm} \vskip-.5cm 
$s_3\  ({\rm GeV^2})$  
\end{minipage}  
        \caption{ Plot of 
$100 \times \frac{{\rm d} |\Gamma_{\rm cusp}(s_3)|}
{{\rm d} s_3}$ over $\frac{{\rm d} \Gamma (s_3)}{{\rm d} s_3}$
around threshold as a function of  
 $s_3$, $ 4 m_{\pi^0}^2\leq s_3\leq 4(2 m_{\pi^+}^2-  
 m_{\pi^0}^2) $, for the decay $K^+\rightarrow \pi^0\pi^0\pi^+$.}  
\label{fig:cuspeffect}
  \end{figure}

The relative effect of the cusp in  $|A_{00+}|^2$ can be seen 
in Figure \ref{fig:cuspeffect}, where we plot
$\frac{{\rm d}\ |\Gamma_{\rm cusp}(s_3)|}
{{\rm d} s_3}$ over $\frac{{\rm d} \Gamma (s_3)}{{\rm d} s_3}$
using the results in  (\ref{ABscatNLO}).
Here, $\Gamma_{\rm cusp}$ is the contribution of  
$v_\pm(s_3) \Delta_{\rm cusp}(s_1,s_3)$ in (\ref{Dcusp}) to
the total $K^+ \to \pi^0 \pi^0 \pi^+ $ decay rate $\Gamma$, 
\ba
\label{cuspdef}
\frac{{\rm d}
 \Gamma_{\rm cusp}(s_3)} {{\rm d} s_3}
= \frac{1}{N} \, {\displaystyle 
\int^{s_1 {\rm max}}_{s_1 {\rm min}}} 
{\rm d} s_1  \,
v_\pm(s_3) \, \, \Delta_{\rm cusp}(s_1,s_3)\ ,
\ea
where $N=512 \pi^3 m_K^3$,  
and $s_{1 {\rm max}}$ and  $s_{1 {\rm min}}$ can be found  in Eq. (4.4)
of \cite{GPS03}.   
Remember that additional pieces
to (\ref{deltaA}) and (\ref{reA00+})
discussed  after  (\ref{range2})
have to be added  when either $s_{1}$  or $s_2$ is above  
the upper limit given in (\ref{range2}).
 If added, these pieces
  would make $\delta \re \overline A_{00+}$ and therefore 
$\Delta_{\rm cusp}(s_1,s_3)$ to diverge
as $s_{1}$  or $s_2$  approaches $(m_{K^+}-m_{\pi^{0}})^2$.
In the integral above 
we do not include those additional pieces
which were also not included in \cite{CI05}
and leave for a future work the study  of the possible solutions
mentioned  above between (\ref{range2}) and (\ref{reB00+}).
So that,  we include in $\Delta_{\rm cusp}(s_1,s_3)$
the same terms included in \cite{CI05}.
We perform the integrals in $x$ and $y$ 
in (\ref{ABscatNLO}), (\ref{ImA++-}) and 
(\ref{deltaA})  numerically.

\begin{figure}
\begin{minipage}[t]{3cm}\vskip-5.5cm 
$\frac{N}{|C'|^2} \, \frac{d\Gamma_{\rm cusp }/ds_3}{v_\pm(s_3)}
\ ({\rm GeV}^6) $\end{minipage}
\includegraphics[{width=10cm}]{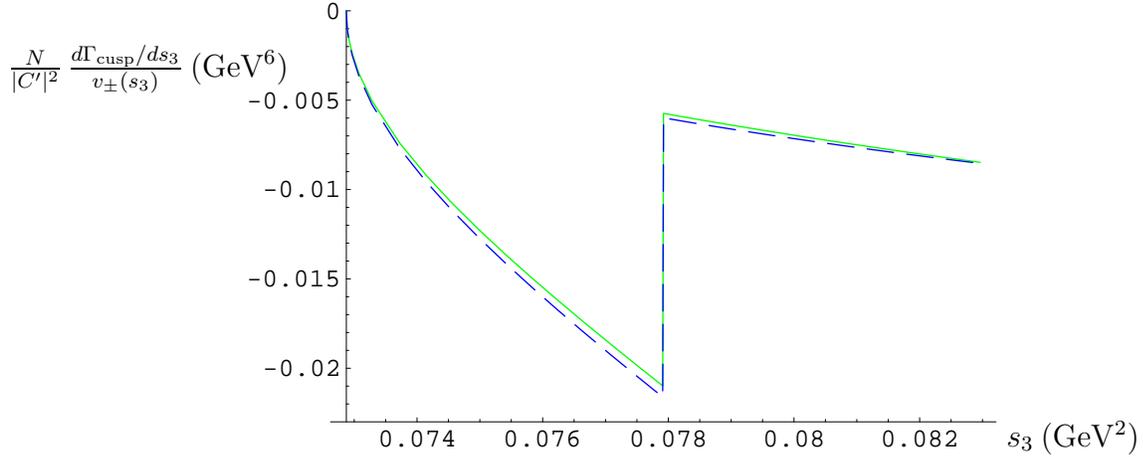}  
\begin{minipage}[t]{1cm} \vskip-.5cm $s_3 \ ({\rm GeV^2})$  
\end{minipage}  
        \caption{ Plot of 
$\frac{N}{|C'|^2} \,  
\frac{d\Gamma_{\rm cusp }/ds_3}{v_\pm(s_3)}$
around threshold as a function of  
 $s_3$, $ 4 m_{\pi^0}^2 \leq s_3 \leq 4 (2 m_{\pi^+}^2-  
 m_{\pi^0}^2) $, for the decay $K^+\rightarrow \pi^0\pi^0\pi^+$.   
The  meaning  of the various lines is explained in the text.}  
\label{fig:cab-ppm}
  \end{figure}

In  Figure \ref{fig:cab-ppm},
we show the cusp effect for $K^+\to\pi^0\pi^0\pi^+$ at NLO.  
 The solid line is our result for 
$\frac{N}{|C'|^2} \, \frac{d\Gamma_{\rm cusp }/ds_3}{v_\pm(s_3)}$
in (\ref{Dcusp}) using (\ref{ABscatNLO}).
Effective $\pi\pi$ scattering lengths near threshold
in (\ref{scat}) are unknowns to be 
fixed by fitting the cusp  effect, however  
for the numerical comparison we take 
them as inputs given by  CHPT \cite{CGL00} at NLO. 
For the rest of inputs 
needed for $\rm Re \overline A_{00+}$  and 
$\rm Re \overline A_{++-}$  we use the
ones in Appendix \ref{inputs}, which  
have been obtained from  a fit to data. 
 
Using the same value of the $\pi\pi$ scattering lengths,
we  also plot   $\frac{N}{|C'|^2} \, 
\frac{d\Gamma_{\rm cusp }/ds_3}{v_\pm(s_3)}$ 
in Figure \ref{fig:cab-ppm} 
as obtained  using  the results in \cite{CI05} 
--this is the dashed curve.
For the   slopes needed for  $\re A_{00+}$  
and $\re A_{++-}$  in  (4.6) and (4.7) in  
\cite{CI05},  we use the  
Taylor expansion of   our NLO in CHPT results
for $\re  A_{00+}$   and $\re A_{++-}$. This makes
the comparison with the solid curve clearer.
 The effect of using instead the slopes   from Taylor expanding  
$|\re  A_{00+}|^2 + |\im A_{00+}|^2$
and $|\re  A_{++-}|^2 + |\im A_{++-}|^2$, respectively,
amounts to a further increase of
  the difference between solid and dashed curves by
around 2 \% below  threshold and    1 \% above threshold.  
Notice that from a fit to data one can just access the 
slopes from $|\re  A_{00+}|^2 + |\im A_{00+}|^2$
and $|\re  A_{++-}|^2 + |\im A_{++-}|^2$.

If one compares the dashed and  the solid curves in Figure
\ref{fig:cab-ppm},  one gets differences below threshold
around 3 \%  while above threshold vary between 2.5 \% and 1 \%.
 They  come from several   order 1\% approximations 
done in \cite{CI05} which we have identified and enumerate below.

The piece $\im \overline A_{00+} \, \re \overline B_{00+}$,
--which we include in the numerics and it was not in  \cite{CI05}--
is nominally order $a^3$   and  
contributes to the cusp below threshold by less than 1 \%.
Notice that this term contains a suppressing velocity factor 
$v_{00}(s_3)$. 
Using a quadratic polynomial in $s_3$  as in \cite{CI05} instead
of the exact NLO CHPT formula is a very good approximation
and  gives  differences  smaller than 0.5 \%.

As said above, the difference between doing  an integral 
in $x$ or  $y$  in (\ref{ABscatNLO}), (\ref{ImA++-}) and 
(\ref{deltaA})  numerically  and doing the linear approximation 
is around  1 \% each.  Since
the cusp  formula in (\ref{Dcusp}) is quadratic 
some of these differences  are actually doubled. 
 At the end of the day, these individually  negligible approximations
produce  the final  differences between dashed and solid
curves quoted above and seen in  Figure \ref{fig:cab-ppm}.
Notice that we don't get the final  difference by summing the
individual ones, it just happens that they  go  in the same
direction.

This provides a first handle on the theoretical   
uncertainty  of the Cabibbo's approach to obtain the
scattering lengths  since  the differences 
quoted above do  not affect the  treatment of $\pi\pi$ scattering.  

Let us now estimate the theoretical uncertainty 
in our approach of determining the effective scattering length
 $a_x(s_3)$ in (\ref{scat}). 
There   are two main sources of theoretical uncertainties.
One is   how good  
is the fit of NLO CHPT formulas to experimental data.   
In case of using CHPT NLO formulas for $\re \, \overline A_{00+}$,
this fit produces central values which agree  
with experiment  within $3\%$ accuracy \cite{GPS03,BDP02}.   
This global fit was done  using  total decay rates and  
Dalitz variable slopes. We want to remark that 
this  fit of  $\re \, \overline A_{00+}$ to data
has to be good enough also away from threshold. Notice that,
for instance, 
  in (\ref{reB00+}) one needs  $\re \, \overline A_{00+}$ 
evaluated at $a(s_3)$ which is typically around
$(m_K^2-m_\pi^2)/2$.
  
The other main  source of uncertainty,
 which is more difficult to estimate   
and  that is  present in  any approach which describes the cusp 
effect to NLO, is the NNLO corrections.  
The only definite way to know it is to calculate these corrections.  
In our approach, this means to calculate 
the real part of $K^+\to \pi^0\pi^0 \pi^+$ at two-loop  
in the isospin conserving  limit   
and make the same analysis that we have done  but at NNLO. 
Notice that in this analysis the two-pion phase space factors 
are the physical ones in order to describe the cusp effect.

Meanwhile, we can just make the following  estimate.  
Going from one order to the next one  in our approach  
implies  that new topologies with an extra $\pi\pi$ scattering   
vertex are needed. For instance, topology C in Figure \ref{topologies}  
appears when going  from LO to NLO, and analogously 
there appear new topologies  when going from  NLO to NNLO. 
Following this line,  a naive estimate of NNLO re-scattering  
effects in $K \to 3 \pi$  is that they are  suppressed with 
respect to LO  by $a_i^2$. Notice that the velocity factors that 
appear after applying  the unitarity  cuts 
can be order one  --see for instance (\ref{reB00+}), where
$v_{\pm}(a(4 m_\pi^2))\simeq 0.6$--
and  do not suppress the naive  $a_i^2$ estimate.
Our estimate  coincides numerically with the one made in  \cite{CI05},
i.e. it gives  around 5 \% corrections from NNLO contributions.
As said above, at NNLO it is possible to   follow a procedure analogous to 
the one  we use  here   to get a more accurate  
measurement of $a_0-a_2$ and check the estimated NNLO uncertainty.  
   
If the theoretical uncertainty from the fit is added to the 5 \% 
of canonical uncertainty assigned
to NNLO   we get a theoretical uncertainty
in our approach of obtaining  the scattering 
lengths from the cusp effect in $K^+ \to \pi^0 \pi^0\pi^+$
between somewhat larger than 5 \% (if uncertaintities are
added quadratically) and 7 \% (if added linearly), i.e.,
we essentially agree therefore with the uncertainty  
quoted by \cite{CI05}.

A further source of error in the final determination  
of the scattering length $a_0-a_2$ from a fit to the cusp effect below   
threshold is due,  as already discussed in \cite{CI05},  
  to the existence  of different strategies to do the fit.  
The three basic  ones are:    
\begin{itemize}  
\item One can consider all the $a_i$ as free  
 parameters in (\ref{ABscatNLO}).   
\item All $a_i$ are fixed to their standard values except the    
combination $a_0-a_2$  in $a_x$ that is extracted from the fit.  
\item One can use CHPT as much as possible. Since the   
combination $a_0-a_2$, or equivalently   
$a_x(s_3)$, only appears in $\overline B_{00+}$,  which   
 is proportional  to the cusp,   
it is enough to keep $\overline B_{00+}$   
in terms of  the scattering lengths. So that, one could
fully use the CHPT predictions   
at NLO in \cite{GPS03,BDP02} for Im $\overline A_{00+}$.  
\end{itemize}  
The comparison of the results for  $a_x$ obtained from  
these different fit strategies  will provide us with another handle  
to estimate the accuracy of the method.  This uncertainty   
will become clear once the fits that we propose are done  with data   
and  to be added to the previous ones.
To the uncertainties discussed above, 
 one  still has to  add the one here from the different 
data fitting strategies.  

\subsection{Cabibbo's Proposal for  Neutral Kaons}  
\label{neutral}  
  In the neutral kaon channel it is also possible   
to measure the scattering lengths combination   
$a_0-a_2$ from the cusp effect in the energy spectrum 
of two $\pi^0$  of $K_L\to\pi^0\pi^0\pi^0$.  
In this case the Bose symmetry of the 
three neutral pions implies that the amplitude  is completely 
symmetric for the interchanges 
$s_1\leftrightarrow s_2\leftrightarrow s_3$.
The amplitude near $\pi^+ \pi^-$ threshold can be written as
\cite{CI05}
\ba
A_{000}&=&\left\{ \begin{array}{ll}
\sum\limits_{i=1,2,3} \left[\overline{A}_{000}(s_i)+ v_\pm(s_i) 
\overline{B}_{000}(s_i)\right]\ ,&  {\rm if\, \,  all \, \,} 
s_i>4 m_{\pi^+}^2 \ ; \\  \\
\left\{ \sum\limits_{i=1,2,3}  \overline{A}_{000}(s_i)\right\} +
 \, i \, v_\pm(s_k) \overline{B}_{000}(s_k) &\\
+\sum\limits_{\scriptstyle
\begin{array}{l}\scriptstyle{i=1,2,3}\\
\scriptstyle{i\neq k}\end{array} }  v_\pm(s_i) 
\overline{B}_{000}(s_i)\ , 
& \left\{ \begin{array}{l} \quad {\rm for} \quad 
s_k<4 m_{\pi^+}^2, \\ k=1,2,3 
\\ \quad {\rm for} \quad s_{i\neq k} > 4 m_{\pi^+}^2\ .
\end{array} \right\} \end{array} \right.
\label{eq:a000}
\ea
The  crucial observation in (\ref{eq:a000})  is that if  
the value  of   $s_k$ is  below threshold, 
the other two variables $s_i$ ($i\neq k$) are of order 
$(m_K^2-m_\pi^2 )/2$, so safely
 above threshold. Thus it is kinematically impossible to cross 
the threshold  with all three variables $s_i$ at the same time.

In the region  where $s_3$ is around threshold, 
one can define
\ba
\overline{A}_{000}^\prime(s_1,s_2,s_3)&=& 
\sum_{i=1,2,3} \overline{A}_{000}(s_i)+ 
\sum_{i=1,2} v_\pm(s_i) \overline{B}_{000}(s_i)\ ,\nn \\
{\rm and} \quad \quad
 \overline{B}_{000}^\prime(s_3)&=& \overline{B}_{000}(s_3)\ .
\label{eq:abdef}
\ea
Equations (\ref{tot})-(\ref{Dcusp}) are now valid also  
for the decay of $K_L\rightarrow\pi^0\pi^0\pi^0$ once  
the substitutions $A_{00+}\rightarrow A_{000}$, 
 $\overline{A}_{00+}\rightarrow \overline{A}_{000}^\prime$ and 
 $\overline{B}_{00+}\rightarrow \overline{B}_{000}^\prime$ are done.

 Again, we write these amplitudes  using  
the approximation (\ref{scat}) for   the amplitudes near threshold.   
The amplitudes contain  $\pi\pi$ re-scattering in all channels and
in those channels where
they are  far from threshold we use full NLO CHPT expressions
which, as already  pointed out  in Section \ref{Cabibbo}, 
 is a better approximation than the effective scattering lengths.
 
 The calculation of  
$\overline A_{000}'$ and $\overline B_{000}'$
is completely analogous to the charged kaon case
-- see Section \ref{charged}. Using the tree-level result 
for the real part of $K_L \to \pi^0\pi^0\pi^0$,  one finds  
\ba \label{AB0LO}  
\re{\overline A_{000}^\prime}\vert_{LO}   &=& 
C'\,(\re \, G_8-G_{27})\mkd ,\nonumber \\  
\im{\overline A_{000}^\prime}\vert_{LO} (s_1,s_2,s_3)  &=&
\re{\overline A_{000}^\prime}\vert_{LO}  
\left[ v_{00}(s_3) a_{00}(s_3) +
\, \sum_{i=1,2} v_{00}(s_i) \frac{m_\pi^2}{32 \pi^2 f_\pi^2}  \right]
\nonumber\\&&+ \sum_{i=1,2} v_{\pm}(s_i) \, 
\fr{s_i-m_\pi^2}{16\pi f_\pi^2} \, f_{000}(s_i)\ , \nn \\
\re{\overline B_{000}^\prime}\vert_{LO} (s_3)  &=& 0 \,, \\  
\im{\overline B_{000}^\prime}\vert_{LO} (s_3)  &=&   
2 a_x(s_3) f_{000}(s_3)\ , 
\ea
with
\ba
f_{000}(s)&=& C' \left[
\re \, G_8\left(s-\vpd\right) + \frac{G_{27}}{6(\vkd-\vpd)}
\left( s (9\vkd-24\vpd)-5\vkc\nonumber\nn \right.\right. 
+ \\ &&
\left. +24\vpc+\vpd\vkd\right)\Big] \ .     
\ea  

The meaning of $LO$ here is the same  that in (\ref{ABscatLO}).   
\par
The effect of the charged pion re-scattering appears also at NLO.
At this order we use the   CHPT  one-loop formula
for $\re \overline{A}_{000}'$  fitted to data. We get 
\ba \label{AB0NLO}  
\re{\overline A_{000}^\prime}\vert_{NLO} (s_1,s_2,s_3)  &=& 
\re{\overline A_{000}^\prime}\vert_{LO}
+\Big[ M_0(s_1)+M_0(s_2)+M_0(s_3)\Big]_{\order(p^4)} + \nonumber \\
&&+ \delta \re \overline A_{000}'(s_1,s_2,s_3)\, ,\nonumber \\  
\im{\overline A_{000}^\prime}\vert_{NLO} (s_1,s_2,s_3)  
&=& \im{\overline A_{000}^\prime}\vert_{LO} 
\,+\, v_{00}(s_3)   a_{00}(s_3) \Big[  
 M_{0}(s_3)+  \widetilde{M}_{0}(s_3) \Big]_{\order(p^4)}    +
 \nonumber \\ && +\Big\lbrack {\rm Im}A^{(6,1)}_{W}(s_1) 
+{\rm Im}A^{(6,2)}_{W}(s_1) 
 +{\rm Im}A^{(6,1)}_{\pi}(s_1)+
{\rm    Im}A^{(6,2)}_{\pi}(s_1)+\nonumber\\
&&\hspace*{1.cm}+s_1\leftrightarrow s_2\Big\rbrack   \nn \,, \\  
\re{\overline B_{000}^\prime}\vert_{NLO} (s_3)  &=& 
F_{000} (s_3) \, ,  \nonumber 
\\  \im{\overline B_{000}^\prime}\vert_{NLO} (s_3)  &=& 
\im{\overline B_{000}^\prime}\vert_{LO} \,+\, 2 a_x(s_3)   
\left[   M_{1}(s_3)+ \widetilde{M}_{2}(s_3)+ 
\right. \nonumber \\ &&
\left.   + \widetilde{M}_{3}(s_3) (m_K^2  
+3 m_\pi^2-2 s_3) -  
\widetilde{M}_{3}^s(s_3)\right]_{\order(p^4)} \, ,  \quad
\nonumber 
\ea  
where we neglected the contributions of  three-pion cut graphs  
which have been shown to be very small in Appendix \ref{threepion}.
The function $F_{000}(s)$ has the expression
\ba
F_{000}(s) &=& 
-a_{00}(s) v_{00}(s) 
\, \im \overline B_{000}(s)   \nonumber\\
&&-  2 a_x(s)^2 v_{00}(s) 
\, \Big\{ \re{\overline A_{000}} \left(s\right)
+ {\displaystyle \int_{-1}^1} \,
{\rm d} x  \, \re{\overline A_{000}}  \left(t^+(s,x)\right) 
 \Big\}  \,   \\ && - a_x(s)   
{\displaystyle\int_{-1}^1} \,{\rm d} x  \,
\im  \tilde A_{+-0}^L ( t^+(s,x), t^-(s,x))\, \, . \nonumber 
\ea
The first two lines here come from diagram B
in Figure \ref{topologies} while the last line
comes solely from diagram C in the same figure.

As already discussed in the case of the analogous calculation
of the discontinuity for the charged $K^+\to \pi^0\pi^0 \pi^+$
decay  in Section \ref{charged}, 
the  integration contour along the real axis $-1 \leq x \leq 1$ 
in $F_{000}(s)$ gives the   correct result just for
small values of $s$, namely \cite{ANI03}, for
\ba
\label{range3}
4 m_{\pi^0}^2  \leq s 
\leq \frac{1}{2} \left(m_{K^0}^2- m_{\pi^0}^2\right) \, . 
\ea
For larger values of $s$ there is an additional piece \cite{ANI03} 
and the same comments and discussion around (\ref{sL}) apply here.

The functions ${\rm Im}A^{(6,i)}_{j}(s)$ 
are  given in Appendix \ref{sec:L000}
and the meaning of $NLO$ in (\ref{AB0NLO})
 is the same  that in (\ref{ABscatNLO}).   
  The imaginary part of the $K_L\to  \pi^+(p_1) \pi^-(p_2) \pi^0(p_3)$ 
decay amplitude,  $\im A_{+-0}^L(t_1,t_2,t_3)$,
 has been obtained here 
using the optical theorem. See Section \ref{FSI} and Appendix 
\ref{sec:L+-0} for the full expression. From that result,  
disregarding the  tiny contribution from the P wave and the pieces  
that do not contribute to singularities, the function 
$\im \tilde A_{+-0}^L(t_1,t_2)$ is 
\ba
\label{ImA+-0}
\im \tilde A_{+-0}^L (t_1,t_2) = 
 a_{+0}(t_1) v_{+0}(t_1)
{\displaystyle\int_{-1}^1} \, {\rm d} y \,
 \re{\overline A_{+-0}^L}(t_1,t^+(t_1,y), t^-(t_1,y))
+ \, t_1 \rightarrow t_2   \,,&&
\ea
where the expression for $\re{\overline A_{+-0}^L}$ at NLO in CHPT can 
be found in \cite{GPS03,BDP02}.  This expression has to be fitted
to  $K_L \to \pi^+\pi^-\pi^0$ data.
Notice that,  in the formula above,  $\pi\pi$ amplitudes
are sometimes evaluated  far from threshold.  
As explained at the introduction  of Section \ref{Cabibbo},  
whenever this happens we use full NLO CHPT 
predictions for the $\pi\pi$ scattering amplitudes 
and not the effective scattering length combinations near
threshold  in (\ref{scat}). 

The contribution $\delta \re {\overline A_{000}'}$ 
comes from the discontinuity across the physical cuts and 
takes into account just the singularities  of $\re {\overline A_{000}'}$
at  $s_i=4 m_{\pi^+}^2$, 
$s_i=(m_{\pi^+}+ m_{\pi^0})^2$  and $s_i=4 m_{\pi^0}^2$
($i=1,2$)   thresholds. These singularities
 start at order $p^6$ in CHPT. We get
\ba
\label{deltaA0}
\delta \re {\overline A_{000}'} (s_1,s_2,s_3) &=&
{\displaystyle \sum_{i=1,2}} 
v_{\pm}(s_i)\,F_{000}(s_i)-
 \, {\displaystyle \sum_{i=1,2,3}} 
 v_{00}(s_i) \, a_{00}(s_i) 
\nonumber  \\  &&
\times 
\, {\displaystyle \int_{-1}^1} \,  {\rm d} x 
\;\Bigg \lbrace \,
 a_{00}(t^+(s_i,x))v_{00}(t^+(s_i,x))
\Big[  
 {\re \overline A_{000}}(t^+(s_i,x)) 
  \nonumber \\   &&  +  {\displaystyle \int_{-1}^1} \, {\rm d} y 
\;   {\re \overline A_{000}} (t^+(t^+(s_i,x),y)) \Big]
+ a_{x}(t^+(s_i,x))v_{\pm}(t^+(s_i,x))\nonumber \\  &&\times
{\displaystyle \int_{-1}^1} \, {\rm d} y  \; 
\re {\overline A_{+-0}} (t^+(t^+(s_i,x),y),t^-(t^+(s_i,x),y),t^+(s_i,x))
\, \Bigg \rbrace \, . 
\ea
 Here, the same comments after (\ref{range2}) apply. I.e., 
the integration contour $-1\leq x \leq 1$ 
  gives the correct result just when 
  both  $s_3$, $s_1$ and $s_2$ are 
all in the range  in (\ref{range3})
--remember that $s_3$ is always around threshold.
For values of $s_1$ or  $s_2$ outside (\ref{range3})
the same discussion after (\ref{range2}) applies.

  In \cite{CI05},  they used the approximation that
 the integrands in (\ref{AB0NLO}), (\ref{ImA+-0}) and
(\ref{deltaA0})  are to a good accuracy linear in the variables
  $x$ and $y$ .  If we make such approximation, we get
\ba
\label{reB000}
\re \overline B_{000}' \vert_{NLO} (s_3)&=& 
- a_{00}(s_3) v_{00}(s_3) \im{\overline B_{000}}(s_3)  \nonumber   \\&&
- 2 \, a_x(s_3)^2 v_{00}(s_3)
\Big\{ \re{\overline A_{000}} \left( s_3\right) +
2 \, \re{\overline A_{000}} \left( a(s_3) \right)   \Big\}  \nonumber  \\
  && - 8 a_x(s_3) a_{+0}(a(s_3)) v_{+0}(a(s_3)) 
\re{\overline A_{+-0}^L}(a(s_3) , a(a(s_3)), a(a(s_3))) 
\\
\label{dreA000}
\delta \re {\overline A_{000}'} (s_1,s_2,s_3) &=& 
- {\displaystyle \sum_{i=1,2}}\,v_\pm(s_i)\,
\Big\{  a_{00}(s_i) v_{00}(s_i) \im{\overline B_{000}}(s_i)  + \nonumber  \\&&
+ 2 \, a_x(s_i)^2 v_{00}(s_i)
\Big\lbrack \re{\overline A_{000}} \left( s_i \right) +
2 \, \re{\overline A_{000}} \left( a(s_i) \right)   \Big\rbrack  
\nonumber  \\
  && + 8 a_x(s_i) a_{+0}(a(s_i)) v_{+0}(a(s_i)) 
\re{\overline A_{+-0}^L}(a(s_i) , a(a(s_i)), a(a(s_i))) \Big\}  
\nonumber \\
&& -2\sum_{i=1,2,3} a_{00}(s_i) v_{00}(s_i) \Big\lbrace
a_{00}\left(a(s_i)\right)v_{00}\left(a(s_i)\right) 
  \left\lbrack \re \overline A_{000}(s_i)
 + 2 \, \re \overline A_{000}\left(a(s_i)\right)
\right\rbrack \nonumber \\ 
&&+ 2 a_x(a(s_i))v_\pm(a(s_i)) \re {\overline A_{+-0}^L}
\left(a\left(a(s_i)\right),a\left(a(s_i)\right),a(s_i)\right)
\Big \rbrace 
\ea
which agree with \cite{CI05}
up to the terms proportional to $v_{00}(a(s_i))$ and 
$v_\pm(a(s_i))$ 
in $\delta \re {\overline A_{000}'}$, which were not included there.
 
\begin{figure}
\begin{minipage}[t]{3cm}\vskip-6.5cm 
$\small{100 \times} \frac{
{\rm d}| \Gamma_{\rm cusp}(s_3)|/{\rm d} s_3}{{\rm d} 
\Gamma(s_3)/{\rm d} s_3}$
 \end{minipage}
\includegraphics[{width=10cm}]{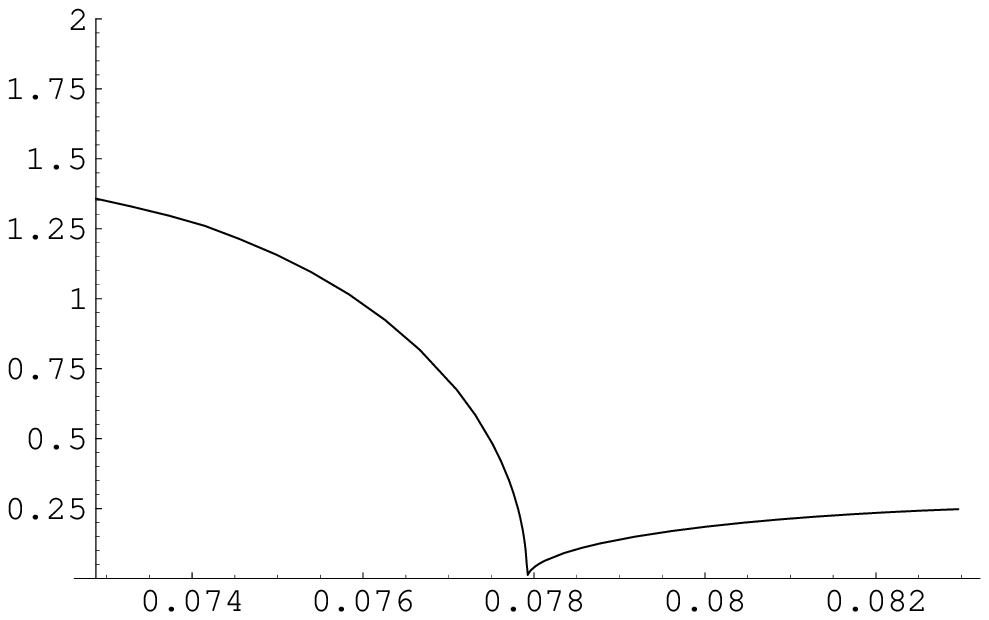}  
\begin{minipage}[t]{1cm} \vskip-.5cm  $s_3 ({\rm GeV}^2)$  
\end{minipage}  
        \caption{ Plot of 
$100 \times \frac{{\rm d} |\Gamma_{\rm cusp}(s_3)|}
{{\rm d} s_3}$ over $\frac{{\rm d} \Gamma (s_3)}{{\rm d} s_3}$
around threshold as a function of  
 $s_3$, $ 4 m_{\pi^0}^2\leq s_3\leq4(2 m_{\pi^+}^2-  
 m_{\pi^0}^2) $, for the decay $K_L\rightarrow \pi^0\pi^0\pi^0$.}  
\label{fig:cuspeffect000}
\end{figure}  

In Figure \ref{fig:cuspeffect000} we show  
the ratio $100 \times \frac{{\rm d} |\Gamma_{\rm cusp}(s_3)|}
{{\rm d} s_3}$ over $\frac{{\rm d} \Gamma (s_3)}{{\rm d} s_3}$  
using the results in  (\ref{AB0NLO})
for $K_L\rightarrow \pi^0\pi^0\pi^0$.
Here, $\Gamma_{\rm cusp}$ is the contribution of  
$v_\pm(s_3) \Delta_{\rm cusp}(s_1,s_3)$ in (\ref{Dcusp}) to
the total $K_L \to \pi^0 \pi^0 \pi^0 $ decay rate $\Gamma$. 
We perform the integrals in $x$ and $y$ 
in (\ref{AB0NLO}), (\ref{ImA+-0}) and 
(\ref{deltaA0})  numerically.

For the sake of numerical comparison with 
the results  in \cite{CI05},  we  do not include the additional pieces 
  discussed after (\ref{range3}) in the integral over $s_1$ 
in (\ref{cuspdef}) defining ${{\rm d} |\Gamma_{\rm cusp}(s_3)|}
/{{\rm d} s_3}$, which were also
not included in that reference either. An analogous  discussion
to the one after (\ref{cuspdef}) applies in this case as well.

\begin{figure}
\begin{minipage}[t]{3cm}\vskip-6.5cm 
$\frac{3 N}{|C'|^2} \, \, 
\fr{d\Gamma_{\rm cusp}/ds_3}{v_\pm(s_3)} ({\rm GeV}^6)
 $\end{minipage}
\includegraphics[{width=10cm}]{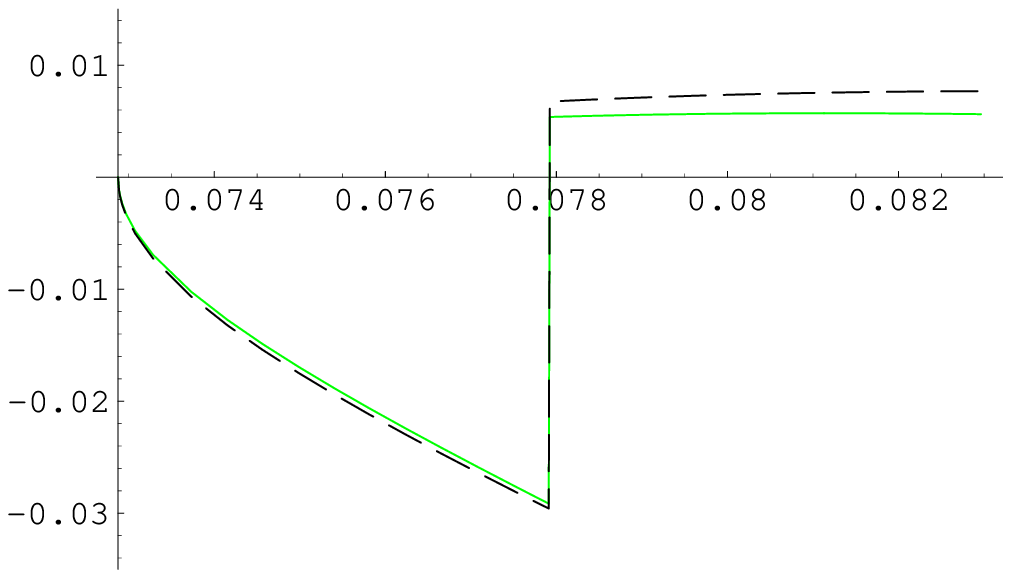}  
\begin{minipage}[t]{1cm}\vskip-.5cm \quad\quad\quad
$\quad\quad\quad s_3 ({\rm GeV^2})$\end{minipage}  
        \caption{ Plot of   
$\frac{3N}{|C'|^2} \, \,
\fr{d\Gamma_{\rm cusp}/ds_3}{v_\pm(s_3)} $   
 around threshold as a function of $s_3$,  
$ 4 m_{\pi^0}^2\leq s_3\leq 4(2 m_{\pi^+}^2- m_{\pi^0}^2) $   
for the decay $K_L\rightarrow \pi^0 \pi^0 \pi^0$.   
The  meaning of the various lines is explained in the text.}  
\label{fig:cab-000}  
\end{figure}

In Figure \ref{fig:cab-000} we plot the cusp effect  
$ \frac{3 N}{|C'|^2} \, \, 
\fr{d\Gamma_{\rm cusp}/ds_3}{v_\pm(s_3)}$ in (\ref{Dcusp})  
for $K_L \to \pi^0\pi^0\pi^0$ at NLO.
The solid line are our results in (\ref{AB0NLO}). 
We use the same inputs as for the
charged case in the previous section.

 We  also plot in  Figure \ref{fig:cab-000},
 the cusp effect $\frac{3 N}{|C'|^2} \, 
\frac{d\Gamma_{\rm cusp }/ds_3}{v_\pm(s_3)}$ 
using  the results in \cite{CI05} --this is the dashed curve.
For the   slopes needed for  $\re A_{000}^L$  
and $\rm Re A_{+-0}^L$  in  (4.58) and (4.59) in  
\cite{CI05},  we use the 
Taylor expansion of  our results for
 $\re  A_{000}^L$  and $\re  A_{+-0}^L$. This makes
the comparison with the solid curve clearer.
 The numerical differences between solid and dashed curves
increase by around 2 \% if 
 instead we use the slopes   from Taylor expanding   
$|\re  A_{000}^L|^2 + |\im A_{000}^L|^2$
and $|\re  A_{+-0}^L|^2 + |\im A_{+-0}^L|^2$, respectively.

If one compares the dashed  and solid 
curves   below threshold in Figure \ref{fig:cab-000}, 
 one gets differences  between 3 \% and 1 \%.
As for the charged case, we have identified the origin of these
differences: they  come from several   order 1\%  
approximations done in \cite{CI05}. Notice that we don't
add them, it just  happens that they go in the same direction. 
The  part of the singularities
in $\delta \re \overline A_{000}'$ that was not included
in \cite{CI05} --see comment after (\ref{dreA000})--
and that we do include in the solid line contributes 
about  +1.5 \% to this difference.
 The other  approximations are analogous to the  ones already enumerated
for the charged case --see previous section.
For instance, the piece $\im \overline A_{000}' \, \re \overline B'_{000}$ 
--which we include and it was not in  \cite{CI05}--
is nominally order $a^3 \, v_{\pm}(s_3)$  and contributes  
to the cusp below threshold by around 1 \%.
Again, at the end of the day, 
 these individually  negligible approximations
produce  the final  differences between dashed and solid
curves quoted above.

Notice again that  the differences between our approach and
the one in \cite{CI05} do  not affect  
the  treatment of the $\pi\pi$ scattering part.  

 Above threshold,  there are  large 
numerical cancellations between 
$\im \overline A'_{000} \, \im \overline B'_{000}$  and 
$\re \overline A'_{000} \, \re \overline B'_{000}$
in  the corresponding expression for
$\Delta_{\rm cusp}$
 in $K_L \to \pi^0\pi^0\pi^0$ --see (\ref{Dcusp}) for the
analogous  charged case.
As a consequence, one should know
both  the  real and imaginary parts of $ \overline A'_{000}$
and $\overline B'_{000}$  with a precision 
better than   1 \% to predict the cusp   above threshold with an
 uncertainty  around  5 \%.
The observed difference between the two curves 
above threshold in Figure \ref{fig:cab-000} 
is due to the need of this fine tuning and is  not physical.
This fact makes the region above threshold not
 suitable to extract the scattering lengths
in $K_L\to\pi^0\pi^0\pi^0$.

Let us now estimate the theoretical uncertainty 
in our approach of determining the effective scattering length
 $a_x(s_3)$ in (\ref{scat}) from $K_L \to \pi^0 \pi^0 \pi^0$. 
 As for the charged case, one  has the theoretical uncertainty from 
the fitting to $\re \overline A_{000}$ 
which includes the theoretical error which measures 
the accuracy of the  formula used to do the fit.
As said above, this has to be checked once
the real fit is done  but we  believe that
it is   realistic  to assume that this theoretical uncertainty
is   around 2 \%. Again, the accuracy of the data should 
be at the level of a few per cent as for the charged case.

If this  is added to the 5 \% 
of canonical uncertainty assigned
to NNLO   we get a theoretical uncertainty
 for the scattering lengths  combination $a_0-a_2$ from
the cusp effect in $K_L \to \pi^0 \pi^0\pi^0$
 between somewhat larger than 5 \% (if uncertainties are
added quadratically) and 7 \% (if added linearly).
I.e., we predict a similar  theoretical accuracy 
in the extraction of $a_0-a_2$ from 
 neutral and  charged kaon cusp effects.
But notice that in the neutral case, this uncertainty 
just applies to the analysis of the data below threshold.
 As said before,  there are very large numerical cancellations
above threshold which preclude the use of these data.

To the  uncertainties discussed above, 
 one  still has to  add the one from the different 
data fitting strategies as described in Section \ref{charged}.  

\section{Summary and Conclusions}  
\label{SC}  
  
In Section \ref{FSI},  
we have presented the  full FSI phases for all $K\to3\pi$ decays at NLO  
in CHPT, i.e. at ${\cal O} (p^6)$ analytically.  
 The two-pion cut contributions for $K^+ \to 3\pi$  
were already presented in \cite{GPS03}. We complete the
calculation  here with the  
three-pion cut contributions and the full result for  
$K_{L,S}\to3\pi$.  The two-pion cut contributions are given analytically
while  the three-pion cut ones are done numerically and checked
always to be negligible.
We used the techniques already explained in \cite{GPS03}  
which are  based  on perturbative unitarity  and analyticity of CHPT. 
  
 In Section \ref{charged}, we study Cabibbo's proposal    
to measure the scattering lengths combination $a_0-a_2$   from 
the cusp effect in the total $\pi^0\pi^0$ pair energy spectrum  
 in $K^+\to\pi^0\pi^0\pi^+$ \cite{CAB04,CI05}.
To be more specific, ours is 
a variation of the original Cabibbo's proposal that uses NLO CHPT
for the real part of $K\to 3\pi$ vertex instead of the
 quadratic polynomial in $s_3$ approximation used in \cite{CAB04,CI05}
plus analyticity and unitarity. 
We studied also the analogous proposal for $K_L\to\pi^0\pi^0\pi^0$  
in Section \ref{neutral}.  

  Notice that we do not use CHPT to predict
the real part  of $K\to 3 \pi$,  but use its 
exact singularity form at NLO in CHPT to fit it  to data above threshold.
If the two-loop CHPT singularity structure was known 
it could be used in order to take this singularity structure
exactly in $\re \overline A_{00+}$. 
The treatment of $\pi\pi$ scattering near threshold 
is independent of this choice
and we treat it in the same way as in \cite{CI05}.
See the introduction of Section \ref{Cabibbo}.

 The cusp effect originates in the different 
contributions to  $K^+\to \pi^0\pi^0\pi^+$   and
$K_L\to \pi^0\pi^0\pi^0$  
amplitudes above and below threshold of $\pi^+\pi^-$ production
in the $\pi^0\pi^0$ pair invariant energy. 
We obtain these contributions 
using  just analyticity and unitarity, in particular applying
Cutkosky rules and   the optical theorem above and below threshold
to calculate the discontinuity across the physical cut.
  This allows us to separate $\pi\pi$ scattering 
--which we want to measure-- from the rest  of $K^+\to \pi^0\pi^0\pi^+$
or $K_L\to \pi^0\pi^0\pi^0$.

We would like to remark here
that making the same approximations  that were done 
 in \cite{CI05} we fully agree with their analytical results.
In particular, we checked that
the use of the quadratic polynomial in $s_3$ in
\cite{CI05} instead of CHPT formulas at NLO produce negligible differences 
--around 0.5  \%-- in $\Delta_{\rm cusp}$ in (\ref{Dcusp}).

 The real part of the discontinuity
 has a singularity when any of the $s_i$ invariant energy
reaches its pseudo-threshold
at $(m_K-m_{\pi^{(i)}})^2$ as described in \cite{ANI03}. We have
discussed how this  singularity appears in our formulas
for the discontinuity and discussed its effects in 
  the description of the cusp effect using the discontinuity
--see  Section  \ref{charged} and Section \ref{neutral}.   

In particular, we pointed out  
that while the presence of that singularity at pseudo-thresholds
does not affect $\re \overline{B}_{00+}(s_3)$ in
(\ref{ABscatNLO}) and $\re \overline{B}_{000}'(s_3)$ in (\ref{AB0NLO})
when $s_3$ is around threshold, 
one needs to take fully into account its  effects for   
$\delta \re {\overline A_{00+}}$ in (\ref{deltaA})  and (\ref{reA00+})
and $\delta \re {\overline A_{000}'}$  in (\ref{deltaA0}) 
and (\ref{dreA000}) when  $s_1$ or $s_2$ is  above $(m_K^2-m_\pi^2)/2$.
For a possible solution  of this problem   which does not simply use
the discontinuity to describe the cusp effect see  \cite{CGKR05}.  
Another possibility  could be to  use,  instead
of  $\delta \re {\overline A_{00+}}$  
 and $\delta \re {\overline A_{000}'}$, 
 the full-two loop (not available yet) finite relevant pieces
to describe  the singularities at   thresholds  at NLO 
 in $\re {\overline A_{00+}}$   and $\re {\overline A_{000}'}$, 
respectively.
This could be fitted to data since we don't want to predict
$\re {\overline A_{00+}}$  and $\re {\overline A_{000}'}$
 but to obtain the best description of them. 
 We leave a more detailed  study of this problem 
and possible solutions for a future work.  

  In Sections \ref{charged}  and \ref{neutral}
we have also discussed  some  approximations done in \cite{CI05}
and the  numerical differences they induce 
in $\Delta_{\rm cusp}$ in (\ref{Dcusp}).
See Figure \ref{fig:cab-ppm} and Figure \ref{fig:cab-000} and text around.
We have identified them and 
found that though each one of them  is individually negligible
(between 0.5 \% to 1 \%) they produce final differences in the
$\Delta_{\rm cusp}$ around  3 \%. 

 In the same sections, 
the theoretical uncertainties  in the determination of 
  $a_0-a_2$ if one uses our formulas to fit the experimental 
data are discussed.   We concluded that for $K^+\to \pi^0\pi^0\pi^+$, 
 this uncertainty is  somewhat larger than 
5 \% if uncertainties are added quadratically  and 7 \% if added linearly. 
 I.e., we essentially agree with the estimate in \cite{CI05}.
Notice that we get our final theoretical uncertainty
as the sum of several order 1\% to 2\% uncertainties to the canonical
NNLO 5 \% uncertainty. 

For the  case $K_L\to \pi^0\pi^0\pi^0$, we get
 that --if  one  uses  just  data below threshold--
the uncertainty in the determination of
$a_0-a_2$ is  of the same order  as for
$K^+ \to \pi^0 \pi^0 \pi^+$.  Above threshold,  we found  
large numerical cancellations  which preclude from using it.

An expansion in the scattering lengths $a_i$ and Feynman diagrams  
were  used in \cite{CI05} to do the power counting and  
obtain  the cusp effect description of  $K^+\to \pi^0\pi^0\pi^+$
at NLO.  In general, when FSI $\pi\pi$ scattering effects
are included at $n$-th 
order\footnote{$n=1$ order stands  for LO contributions.},
 there  appear new topologies in $K\to 3\pi$ 
--topology C at NLO order, for instance--
which give contributions  to $\Delta_{\rm cusp}$ 
of order  $a_i^{n}$. 
 The canonical uncertainty of the $n$-th order results  is thus $a_i^{n}$.
Notice that the velocity factors that 
appear after applying  the unitarity  cuts 
can be order one  --see for instance (\ref{reB00+}), where
$v_{\pm}(a(4 m_\pi^2))\simeq 0.6$--
and  do not suppress the naive  $a_i^n$ estimate.

Our estimate  for the uncertainty from NNLO, $\sim a_i^2$,  
coincides numerically with the one made 
in  \cite{CI05},i.e. it gives  around 5 \%.
We conclude that one cannot expect to decrease this canonical $5\%$
 theoretical uncertainty  of the NLO result 
unless one includes  $\pi\pi$ scattering effects at NNLO. 
If one wants  to  
reach the per cent level in the uncertainty of the  
determination of $a_0-a_2$ from the cusp effect,   
one would need  to include those  NNLO   re-scattering   effects.  
As said above, at NNLO it is possible to   follow a procedure analogous to 
the one  we use  here   to get a more accurate  
measurement of $a_0-a_2$ and check the estimated NNLO uncertainty.  

We have just included isospin breaking due to the different thresholds 
using   two-pion physical phase spaces in the optical theorem
and Cutkosky rules.
This is needed to describe the cusp effect.
The rest of NLO isospin breaking is expected to be  important  just 
at NNLO. At that order,  isospin breaking effects in $\pi\pi$ 
scattering at threshold --both from quark masses
and from electromagnetism--  will have to be implemented 
and their  uncertainties added.  
  
 Finally, we believe that it is interesting to continue investigating
the proposal in \cite{CAB04,CI05} to measure
the non-perturbative $\pi\pi$ scattering lengths from the cusp
effect in $K^+\to \pi^0 \pi^0\pi^+$ and 
$K_L\to \pi^0 \pi^0\pi^0$. Another interesting direction is to develop
an effective field theory in the scattering lengths which could both
check the results in \cite{CI05} and allow to go to NNLO.
 This type of studies is already underway and firsts results were
presented \cite{GAS06,CGKR05}.

\section*{Acknowledgments}  
It is a pleasure to acknowledge useful discussions    
with Hans Bijnens, J\"urg Gasser,  Gino Isidori and Toni Pich.
This work has been supported in part by  European Commission (EC)  
 RTN Network EURIDICE Contract No. HPRN-CT2002-00311 (J.P. and I.S.),
the HadronPhysics I3 Project (EC) Contract No. 
RII3-CT-2004-506078 (E.G.), 
 by  MEC (Spain)  and FEDER (EC)  Grant Nos. FPA2003-09298-C02-01 (J.P.)  
 and  FPA2004-00996 (I.S.), and by Junta de Andaluc\'{\i}a    
Grants No. P05-FQM-101 (E.G. and J.P.) and P05-FQM-347 (J.P.).
 E.G. is indebted to the EC 
 for the  Marie Curie  Fellowship No. MEIF-CT-2003-501309.
I.S. wants to thank the Departament d'ECM, Facultat de F\'{\i}sica, 
Universitat de Barcelona (Spain) for kind hospitality.
  
\section*{Appendices}  
  
\appendix  
\section{Numerical Inputs}  
\label{inputs}  
  
 Here we discuss the numerical inputs we use in  the numerical  
applications. In \cite{BDP02}, a fit to all available $K\to \pi\pi$   
amplitudes at NLO in CHPT and $K \to 3\pi$ amplitudes and  
slopes in $K\to 3\pi$ amplitudes was done. This was done in the   
isospin limit. More recently, this fit was updated in \cite{BB05}  
with  new data on slopes and including also the full isospin breaking  
 effects. Though our calculation of FSI at NLO uses  
 isospin limit results we will use the results obtained in this last fit.  
The reason is that  the change in the fit results  is due to both the   
new data used and the isospin breaking corrections at the same level  
and therefore cannot be disentangled.  In addition, the main isospin   
breaking effects due to the kinematical factors is also taken into   
account in our results.  
\begin{table}  
\begin{center}
\label{tabKvalues}  
\begin{tabular}{||c|c||}\hline  
 & $\re \widetilde K_i(M_\rho)$ from \cite{BB05}\\    
\hline\hline  
$\widetilde K_2(M_\rho)$ & $G_8 \times (48.5\pm 2.4) \cdot 10^{-3}$ \\   
\hline  
$\widetilde K_3(M_\rho)$ & $G_8 \times (2.6 \pm 1.2) \cdot 10^{-3}$ \\  
\hline  
$\widetilde K_5(M_\rho)$ & $- G_{27} \times
(41.2\pm 16.9) \cdot 10^{-3}$\\  
\hline  
$\widetilde K_6(M_\rho)$ & $- G_{27} \times 
(102 \pm 105) \cdot 10^{-3}$\\\hline  
$\widetilde K_7(M_\rho)$ & $G_{27} \times    
(78.6\pm 33) \cdot 10^{-3}$ \\\hline  
\end{tabular}  
\caption{Results for the  order $p^4$ counterterms
$\re \, \widetilde K_i$ from the fit  to data done in \cite{BB05}.   
The values of $\re \, \widetilde K_i$ which do not   
appear are zero. For definitions of the counterterms, 
see \cite{GPS03,BDP02}.}   \end{center}\end{table}  
    
 At order $p^4$, NLO in  CHPT, the results in \cite{BB05} are equivalent  
[using $F_0=87.7$ MeV] to  
\ba  
\re \, G_8 = 6.6 \pm 1.1 \hspace*{0.5cm}  
{\rm and} \hspace*{0.5cm}
 G_{27} = 0.44 \pm 0.09 \, .     
\ea  
In this normalization, $\re \, G_8 = 1 = G_{27}$ at large $N_c$.  
($N_c$ is the number of colors of QCD).   
  
 For the NLO prediction of the FSI, we only need the real part  
of the counterterms and in particular the combinations in Table 1  
 of \cite{GPS03}  
which from the new fit in \cite{BB05} are given in   
Table \ref{tabKvalues}. These were obtained from a fit of the NLO  
in CHPT results to experimental  data.  Since we only want 
a good fit to data, one can fix them to any of these values
which produce equally good fits.
   
\section{FSI Phases at NLO for Neutral Kaon Decays: Two-Pion Cuts}  
\label{twopion}  
  
Here we include the analytical results for the two-pion  
cuts contributions to the  dispersive part   
of the neutral decay amplitudes at NLO in CHPT, i.e. $\order(p^6)$, 
coming   from diagrams  in Figure \ref{fig:000}-\ref{fig:+-0}. 
Analogous results   
for the charged decay amplitudes as well as a more detailed description   
of the method, based on the use of the optical theorem,    
can be found in Appendix E of the first reference in \cite{GPS03}.   
The fully NLO FSI phases are completed by the   
calculation of the three-pion cut contributions,  
 whose analytical results are given in   
Appendix \ref{threepion}.  
In Subsections \ref{sec:L000}, \ref{sec:L+-0} and  
 \ref{sec:S+-0} we give the dispersive part of  
the amplitudes  $A_i$ for $K_{L,S} \to 3\pi$ denoted by  
Im \, $A^6_i$  where used the super-index  
$6$ to indicate the CHPT order.  
  
\subsection{Notation}  
\label{sec:not6}  
In all the definitions and results written in this Appendix and in 
Appendices \ref{sec:L000}, \ref{sec:L+-0} and \ref{sec:S+-0}, 
the functions $M_i(t)$ and $P_i(t)$ are only 
the real part of the corresponding functions 
defined in \cite{BDP02} and \cite{BKM91} respectively. The 
expressions in Eqs. (\ref{im0001})-(\ref{im0002}) are thus real. 

We define  
\ba  
\widetilde{M}_i(s)&=& \int_{-1}^1 {\rm d} x \;    
\left. M_i(a(s)+ b(s) x )   \right|_{p^4} \, ,\nonumber\\  
\widetilde{M}_i^s(s)&=&   
\int_{-1}^1 {\rm d}  x\;  \left.  (a(s)+b(s) x ) \,   
M_i(a(s)+b(s) x )\right|_{p^4}\label{tildeMis} \, ,\nonumber \\ 
\widetilde{M}_i^{ss}(s)&=&   
\int_{-1}^1 {\rm d} x \; \left.  
(a(s)+b(s) x)^2 M_i(a(s)+b(s) x)\right|_{p^4} \, , \\  
\label{def_a_b}
{\rm with} \hspace*{1cm} 
a(s)&=& \fr{1}{2}( 3 s_0-s) 
+ \frac{(m_K^2 - m_{\pi^{(3)}}^2 ) 
(m_{\pi^{(1)}}^2 -m_{\pi^{(2)}}^2)}{2 s },\nn\\  
{\rm and} \hspace*{1cm} b(s)&=& 
\frac{1}{2 s} \Big\{ 
\Big[(s-(m_{\pi^{(1)}}+m_{\pi^{(2)}})^2)
(s- (m_{\pi^{(1)}}-m_{\pi_{(2)}})^2)\Big]       \times \nn \\ &&\times
 \Big[(s-(m_K+m_{\pi^{(3)}})^2)
(s- (m_K-m_{\pi^{(3)}})^2)\Big]  \Big\}^{1/2}\nn  \\
  \ea  
for $s=(k-p_3)^2$, see (\ref{defdecays})
for definition of these momenta.

The amplitudes at ${\cal O}(p^4)$ for the $\pi\pi\rightarrow \pi\pi$
 scattering  in a theory with  
three flavors can be found  in \cite{BKM91}.  
We decompose the amplitudes in the various cases as follows.  
For the case $\pi^+\pi^+\rightarrow \pi^+\pi^+$   
the amplitude at ${\cal O}(p^4)$  
is  
\ba  
\Pi_1&=& P_1(s)+P_2(s,t)+P_2(s,u) .  
\ea  
For the case $\pi^0\pi^0\rightarrow \pi^+\pi^-$ the amplitude 
at ${\cal O}(p^4)$  
is  
\ba  
\Pi_2&=& P_3(s)+P_4(s,t)+P_4(s,u) .  
\ea  
For the case $\pi^+\pi^-\rightarrow \pi^+\pi^-$ 
the amplitude at ${\cal O}(p^4)$  
is  
\ba  
\Pi_3&=& P_5(s)+P_6(s,t)+P_6(s,u)+P_7(s,t)-P_7(s,u) . \nn \\ 
\ea  
Finally, the amplitude  
$\pi^0\pi^0\rightarrow \pi^0\pi^0$ at ${\cal O}(p^4)$  
is  
\ba  
\Pi_4&=& P_8(s)+P_8(t)+P_8(u) .  
\ea  
The value for the various $P_i$ can be obtained from \cite{BKM91}.  
In the following we use 
\ba\!\!\!\!\!\!\!\!\!  
\widetilde{P}^{(n,m)}_i(s)&=& s^n     c(s)^m \, 
\int_{-1}^1 {\rm d} x\; \, (1- x )^m   \,  
P_i(s,c(s) (1- x)),\nn \\  
\!\!\!\!\!\!\!\!\!  
\hat{P}^{(n)}_{1,i}(s)&=& c(s)^n \,
\int_{-1}^1 {\rm d} x\; (1-x)^n \,   
P_i(c(s)(1+x),c(s)(1-x)),\nn \\  
\!\!\!\!\!\!\!\!\!  
\hat{P}^{(n)}_{2,i}(s)&=& 
c(s)^n \int_{-1}^1 {\rm d} x \;
(1- x)^n   P_i(c(s)(1- x ),s) ,\\  
\!\!\!\!\!\!\!\!\!  
{\rm with} \hspace*{1cm} 
c(s)&=&\fr{1}{2} (4 m_\pi^2-s)  .
\ea  
\subsection{FSI for $K_L\rightarrow \pi^0\pi^0\pi^0$ at NLO}  
\label{sec:L000}

\begin{figure}
\begin{center}
\includegraphics[{width=10cm}]{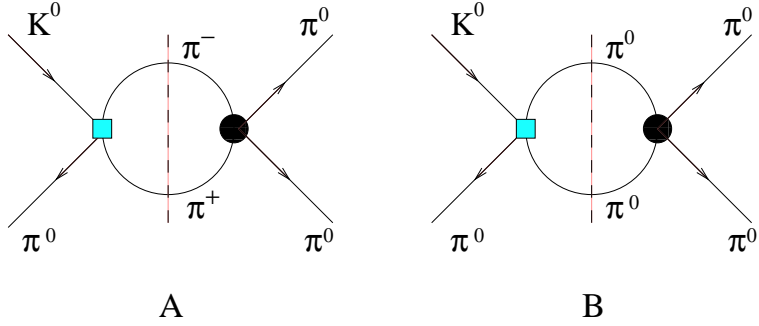}  
\end{center}
        \caption{Diagrams for the calculation of FSI for  
          $K_L\rightarrow\pi^0\pi^0\pi^0$ using the optical theorem.  
The square is the weak vertex and the circle is the strong one.}  
\label{fig:000}
\end{figure}  
  
Diagrams  A --two charged pions in the loop--  
  and B --two neutral pions in the loop-- in Figure \ref{fig:000}  
correspond to the two possible contributions for  
$K_L \to \pi^0\pi^0\pi^0$.  
We first compute the case when the weak vertex   
in Figure \ref{fig:000} is of $\order(p^4)$ and the   
strong vertex of $\order(p^2)$. The results are   
\ba  \label{im0001}
{\rm Im}A^{(6,1)}_{W}(s_1,s_2,s_3)&=&{\rm Im}A^{(6,1)}_{W}(s_1)+{\rm  
  Im}A^{(6,1)}_{W}(s_2)+{\rm Im}A^{(6,1)}_{W}(s_3)\,,  
\nonumber\\  
{\rm Im}A^{(6,1)}_{W}(s) &=& \fr{\s(s)}{16\pi f_\pi^2}(s- m_\pi^2)\left[  
 M_{1}(s)+ \widetilde{M}_{2}(s)+ \widetilde{M}_{3}(s) (m_K^2  
+3 m_\pi^2-2 s)  \right.   
-\left.\widetilde{M}_{3}^s(s)\right]_{\order(p^4)}  \nonumber
\\  
{\rm Im}A^{(6,2)}_{W}(s_1,s_2,s_3)&=&  
{\rm Im}A^{(6,2)}_{W}(s_1)+{\rm  
  Im}A^{(6,2)}_{W}(s_2)+{\rm Im}A^{(6,2)}_{W}(s_3)\,,  
\nonumber \\  
{\rm Im}A^{(6,2)}_{W}(s)  
&=&  
\fr{\s(s)}{32\pi f_\pi^2} m_\pi^2\left[  
 M_{0}(s)+  \widetilde{M}_{0}(s) \right]_{\order(p^4)}.  
\ea  
Then we  consider the same diagrams of Figure   
\ref{fig:000} with a weak vertex of $\order(p^2)$ and a strong   
vertex of $\order(p^4)$. We have  
\ba  
{\rm Im}A^{(6,1)}_{\pi}(s_1,s_2,s_3)&=&{\rm Im}A^{(6,1)}_{\pi}(s_1)+{\rm  
  Im}A^{(6,1)}_{\pi}(s_2)+{\rm Im}A^{(6,1)}_{\pi}(s_3)\,,  
\nonumber \\  
{\rm Im}A^{(6,2)}_{\pi}(s)  
&=& \fr{\s(s)}{16\pi}  \left. (M_{1}(s)+  
M_{3}(s) (m_K^2+m_\pi^2-3 s))  
\right|_{p^2}\   
(P_3(s)+\widetilde{P}^{(0,0)}_4(s)) \ ,  
\nonumber \\   
{\rm Im}A^{(6,2)}_{\pi}(s_1,s_2,s_3)&=&{\rm Im}A^{(6,2)}_{\pi}(s_1)+{\rm  
  Im}A^{(6,2)}_{\pi}(s_2)+{\rm Im}A^{(6,2)}_{\pi}(s_3)\,,\nonumber \\  
{\rm Im}A^{(6,2)}_{\pi}(s)  
&=& \fr{\s(s)}{32\pi}  \left. (M_{0}(s_3)+  
M_{0}(s_1)+M_{0}(s_2))  
\right|_{p^2}\   
(P_8(s)+\widetilde{P}^{(0,0)}_8(s))  
\,,  
\ea  
for those diagrams with two charged and two neutral pions in the loop   
respectively.    
  
\subsection{FSI for $K_L\rightarrow\pi^+\pi^-\pi^0$ at NLO}  
\label{sec:L+-0}

\begin{figure}
\begin{center}
\includegraphics[{width=13cm}]{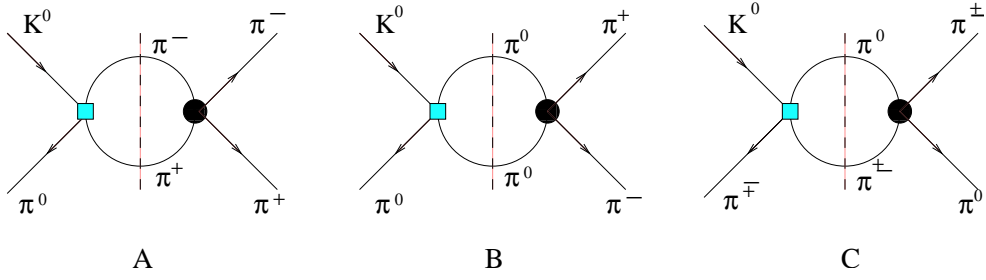}  
\end{center}
        \caption{Diagrams for the calculation of FSI for  
          $K_L\rightarrow\pi^+\pi^-\pi^0$ using the optical theorem.   
The square is the weak vertex and the circle is the strong one.}  
\label{fig:+-0}  
\end{figure}  
  
The calculation is analogous to the one for   
$K_L\rightarrow \pi^0\pi^0\pi^0$. The three contributions in   
Figure \ref{fig:+-0} correspond to two charged pions   
in the loop --results in Eqs. (\ref{case1}) and (\ref{case1p}),  
two neutral pions  in the loop --results in Eqs.   
(\ref{case2}) and (\ref{case2p})-- and loops with one neutral   
and  one charged pions. In   
the last case we have a S-wave contribution   
--results in (\ref{case3S}) and (\ref{case3Sp}), and a   
P wave contribution  --results in (\ref{case3P})  
and (\ref{case3Pp}), respectively.  
Eqs.  (\ref{case1}), (\ref{case2}), (\ref{case3S}),   
and (\ref{case3P}) are the results of using   
the weak vertex at ${\cal{O}}(p^4)$  and the strong vertex at  
${\cal{O}}(p^2)$.  
Eqs.  (\ref{case1p}), (\ref{case2p}), (\ref{case3Sp}), and   
(\ref{case3Pp})  are the results of using   
the weak vertex at ${\cal{O}}(p^2)$  and the strong vertex at  
${\cal{O}}(p^4)$.  
\ba\label{case1}  
{\rm Im}A^{(6,1)}_W&=& \fr{\s(s_3)}{16\pi f_\pi^2}(s_3)\left[  
 M_{1}(s_3)+ \widetilde{M}_{2}(s_3)+ \widetilde{M}_{3}(s_3) (m_K^2  
+3 m_\pi^2-2 s_3)  \right.   
\nonumber\\  
&&  
\left.-\widetilde{M}_{3}^s (s_3)\right]_{\order(p^4)}\,,\\  
\label{case2}  
{\rm Im}A^{(6,2)}_W&=& \fr{\s(s_3)}{32\pi f_\pi^2} (s_3-m_\pi^2)\left[  
 M_{0}(s_3)+  \widetilde{M}_{0}(s_3) \right]_{\order(p^4)}\,,\\  
\label{case3S}  
{\rm Im}A^{(6,3)}_{W,S}&=& \fr{\s (s_1)}{32\pi f_\pi^2}(m_\pi^2-s_1/2)\left[  
 2 M_{2}(s_1)+    
\widetilde{M}_{2}(s_1)+ \widetilde{M}_{1}(s_1) \right.   
\nonumber \\   && 
-\widetilde{M}_{3}(s_1) (m_K^2  
+3 m_\pi^2-2 s_1)   
\left.+\widetilde{M}_{3}^s (s_1)\right]_{\order(p^4)} + 
s_1\leftrightarrow s_2\,,\\  
\label{case3P}  
{\rm Im}A^{(6,3)}_{W,P}&=&\fr{\s(s_1)}{64\pi f_\pi^2}  
 \fr{s_1(s_2-s_3)}{s_1^2-2 s_1  
 (m_K^2+m_\pi^2)+(m_K^2-m_\pi^2)^2}  
\nonumber\\ && \times  
\left[  
(s_1 -m_K^2-3 m_\pi^2)(\widetilde{M}_2(s_1)-\widetilde{M}_1(s_1)  
\right.  
\nonumber \\ &&  
\left.  
+(2  
 s_1-m_K^2-3 m_\pi^2)\widetilde{M}_3(s_1))+  
(2 \widetilde{M}_2^s(s_1)-2 \widetilde{M}_1^s(s_1)  
\right.  
\nonumber \\ &&  
+\left.  
(5  
 s_1-3 m_K^2-9 m_\pi^2)\widetilde{M}_3^s(s_1))  
+2 \widetilde{M}_3^{ss}(s_1)  
+\fr{8}{3} b(s_1)^2 M_3(s_1)   
\right]_{\order(p^4)}  
 \nonumber \\ &&  
+s_1\leftrightarrow s_2\,.  
\\  
\label{case1p}  
{\rm Im}A^{(6,1)}_\pi &=& \fr{\s(s_3)}{16\pi}    
\left. (M_{1}(s_3)+  
M_{3}(s_3) (m_K^2+3 m_\pi^2-3 s_3) )  
\right|_{p^2}\   
( P_5(s_3)+ \widetilde{P}^{(0,0)}_6(s_3) )\, ,\\ 
\label{case2p}  
{\rm Im}A^{(6,2)}_\pi&=& \fr{\s(s_3)}{32\pi}  \left. (M_{0}(s_3)+  
M_{0}(s_1)+M_{0}(s_2))  
\right|_{p^2}\   
(P_3(s_3)+\widetilde{P}^{(0,0)}_4(s_3))\,,\\  
\label{case3Sp}  
{\rm Im}A^{(6,3)}_{\pi,S}&=& \fr{\s(s_1)}{32\pi}  \left. \left( M_{1}(s_1)+  
\fr{1}{2}M_{3}(s_2) (s_1-s_3)+\fr{1}{2}M_{3}(s_3) (s_1-s_2)\right)  
\right|_{p^2}  
\nonumber \\  
&&\times  
\Big(\widetilde{P}^{(0,0)}_3(s_1)  
+\hat{P}^{(0)}_{1,4}(s_1)+\widetilde{P}^{(0)}_{2,4}(s_1)\Big)  
+s_1\leftrightarrow s_2  
\\  
\label{case3Pp}  
{\rm Im}A^{(6,3)}_{\pi ,P}&=&  
\fr{\s(s_1)}{64\pi}\left.\left(3 M_{3}(s_1)(s_2-s_3)\right)\right|_{p^2}\fr{1}{s_1-4 m_\pi^2}\nonumber \\ \et  \times 
\Big( (s_1-4 m_\pi^2)  
(\widetilde{P}^{(0,0)}_3(s_1)-\hat{P}^{(0)}_{1,4}(s_1)+  
\widetilde{P}^{(0)}_{2,4}(s_1))+2 \widetilde{P}^{(1,0)}_3(s_1)  
\nonumber \\ \et  
- 2 \widetilde{P}^{(1)}_{1,4}(s_1) +2\hat{P}^{(1)}_{2,4}(s_1)\Big)  
+s_1\leftrightarrow s_2\  .  
\ea  
  
\subsection{FSI for $K_S\rightarrow \pi^+\pi^-\pi^0$ at NLO}  
\label{sec:S+-0}  
  
The diagrams  contributing to this decay are  the same  
 depicted in Figure \ref{fig:+-0}.    
The result corresponding to  diagram A in Figure \ref{fig:+-0}   
 is in Eq. \ref{kscase1}  for the weak vertex  at ${\cal{O}}(p^4)$ and  
the  strong vertex at ${\cal{O}}(p^2)$  and in    
Eq. \ref{kscase1p} for the weak vertex  at ${\cal{O}}(p^2)$ and  the  
strong vertex at ${\cal{O}}(p^4)$.  
  In this case there is only  P wave contribution.  
  
Diagram  B in Figure \ref{fig:+-0} does  not contribute.  
Diagram C in Figure \ref{fig:+-0}   gives  both an S-wave  
-- results in Eqs. (\ref{kscase2S}) and (\ref{kscase2Sp})--  
 and a  P wave contribution -- results in (\ref{kscase2P})  
 and (\ref{kscase2Pp}).  Equations (\ref{kscase2S}) and (\ref{kscase2P})   
are the results of the case  in which the  
 weak vertex is  at ${\cal{O}}(p^4)$ and  
 strong vertex at ${\cal{O}}(p^2)$ in diagram C.  
Equations (\ref{kscase2Sp}) and (\ref{kscase2Pp})   
are the results for the case  in which the weak vertex is    
at ${\cal{O}}(p^2)$ and the strong vertex at ${\cal{O}}(p^4)$  
in diagram C.  
 \ba  
\label{kscase1}  
{\rm Im}A^{(6,1)}_{W ,P}&=&  
\fr{\s(s_3)}{16\pi f_\pi^2}  
\fr{s_3(s_1-s_2)}{s_3^2-2(m_K^2+m_\pi^2)s_3+(m_K^2-m_\pi^2)^2}\left\{  
a(s_3)(\widetilde{M}_4(s_3)+ (2 a(s_3)  
\right.
\nonumber \\ \et\left. -s_3) \widetilde{M}_5(s_3))  
-(\widetilde{M}_4^s(s_3)+ (3 a(s_3)-s_3)  
\widetilde{M}_5^s(s_3))+\widetilde{M}_5^{ss}(s_3)  
\right.\nonumber \\ \et\left.  
-\fr{2}{3} b(s_3)^2 M_6(s_3)\right\}  
\\  
\label{kscase2S}  
{\rm Im}A^{(6,2)}_{W ,S}&=&\fr{\s(s_1)}{64\pi f_\pi^2}   
(2m_\pi^2-s_1)\left\{  
2 M_4(s_1)-  
\widetilde{M}_4(s_1) +(m_K^2+3 m_\pi^2-2 s_1) (\widetilde{M}_5(s_1)  
\right.   
\nonumber\\\et  
\left.  
-\widetilde{M}_6(s_1))-\widetilde{M}_5^s(s_1)+\widetilde{M}_6^s(s_1)  
\right\}-s_1\leftrightarrow s_2\, ,  
\\  
\label{kscase2P}  
{\rm Im}A^{(6,2)}_{W ,P}&=&  
\fr{\sigma(s_1)}{32\pi f_\pi^2}\fr{s_1(s_3-s_2)}{s_1^2-2  
  s_1(m_K^2+m_\pi^2)+(m_K^2-m_\pi^2)}  
\nonumber\\\et \times  
 \biggl\{  
\fr{4}{3} b(s_1)^2 M_ 5(s_1)+ a(s_1) (\widetilde{M}_4(s_1)-(2  
  a(s_1)-s_1) (\widetilde{M}_5(s_1)+\widetilde{M}_6(s_1)))  
\nonumber\\\et  
-\widetilde{M}_4^s(s_1)+  
(3 a(s_1)-s_1) (\widetilde{M}_5^s(s_1)+\widetilde{M}_6^s(s_1))-  
\widetilde{M}_5^{ss}(s_1)-\widetilde{M}_6^{ss}(s_1)  
\biggr\}  
\nonumber\\\et  
-s_1\leftrightarrow s_2  
\\  
\label{kscase1p}  
{\rm Im}A^{(6,1)}_{\pi ,P}&=&-  
\fr{\sigma(s_3)}{16\pi}\fr{\left. (M_4(s_1)-M_4(s_2))\right|_{p^2}  
}{s_3-4 m_\pi^2}\left[(s_3-4 m_\pi^2)\widetilde{P}_7^{(0,0)}(s_3)+2  
  \widetilde{P}_7^{(0,1)}(s_3)\right]\, ,\nonumber \\  
&&\\  
\label{kscase2Sp}  
{\rm Im}A^{(6,2)}_{\pi ,S}&=&  
\fr{\sigma(s_1)}{64 \pi} \left. M_5(s_1)\right|_{p^2}  
(3 s_1- (m_K^2+3 m_\pi^2)) (  
\widetilde{P}_3^{(0,0)}(s_1)+\widetilde{P}_{2,4}^{(0)}(s_1)  
\nonumber \\ \et  
+\widetilde{P}_{1,4}^{(0)}(s_1))-s_1\leftrightarrow  
s_2 \,,\\  
\label{kscase2Pp}  
{\rm Im}A^{(6,2)}_{\pi ,P}&=&  
\fr{\s(s_1)}{64\pi}\left.\left(M_4(s_2)-M_4(s_3)\right)\right|_{p^2}\fr{1}{s_1-4 m_\pi^2}\nonumber \\ \et  
\times \Big( (s_1-4 m_\pi^2)  
(\widetilde{P}^{(0,0)}_3(s_1)-\hat{P}^{(0)}_{1,4}  
(s_1)+\widetilde{P}^{(0)}_{2,4}(s_1))+2 \widetilde{P}^{(1,0)}_3(s_1)  
\nonumber \\ \et  
-2 \widetilde{P}^{(1)}_{1,4}(s_1) +2\hat{P}^{(1)}_{2,4}(s_1)\Big)   
-s_{1}\leftrightarrow s_{2}\,. \label{im0002}
\ea  

\section{Three-Pion Cut Contributions to  FSI Phases}  
\label{threepion}  

\begin{figure}
\begin{center}
\includegraphics[{width=11cm}]{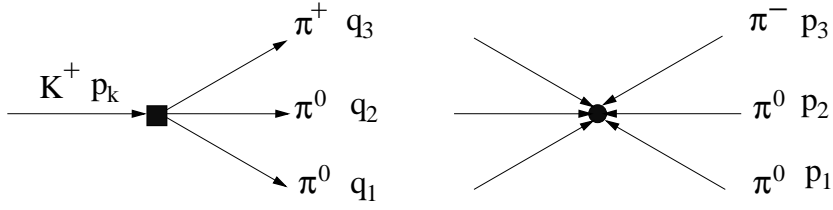}
\end{center}
        \caption{Example of diagram for the calculation of FSI for
 $K^+\to\pi^0\pi^0\pi^+$ with momentum assignment. }
\label{fig:sunset2}
\end{figure}

To calculate the contributions of the topologies 
in Figure \ref{topologies} C, D,  and E to the FSI,
 we  need the tree level  
vertices of $K\to 3\pi$ and $3\pi\to
 3\pi$. The  complete tree level  amplitude of the 
$3\pi\to 3\pi$ scattering can be easily calculated  
using the code {\tt  Ampcalculator} 
developed in \cite{UE:05}. 
In this way all possible  $3\pi\to 3\pi$ are included.
In order to perform the  integral in the phase space as prescribed 
in the optical theorem we assign the momentum to  pions as shown 
in an example in  Figure \ref{fig:sunset2}.
The momentum assignment is done
 preserving the   suffix $3$ for the {\it odd} pion  
in the $K^+\to 3\pi$ vertex.

In order to perform the integral in the phase space over  momentum
 $q_i$ it is necessary the fix a reference frame.
We chose the reference frame in which the decaying kaon is at rest 
 and the momentum $p_3$ defines the z-axis,
\ba
p_K=(M_K,0,0,0), \quad&& p_3=(p_3^0,0,0,\bar p_3), \nonumber\\
  p_1=(p_1^0,0,\bar p_1 \sin\eta,\bar p_1\cos\eta),\quad&&
 p_2=-p_K-p_1-p_3\ . 
\ea 
The momenta $q_i$ describe also a decay of a kaon into three pions. 
The most general expression for these momenta, in the chosen reference
 frame, is
\ba
q_i&=&R_z(\alpha) R_y(\beta)R_z(\gamma)\  r_i\nonumber\ , \\
 r_3=(r_3^0,0,0,\bar r_3), \quad   r_1&=&(r_1^0,0,\bar r_1 
\sin\theta,\bar r_1\cos\theta), \quad 
 r_2=p_K-r_1-r_3\ , 
\ea 
and $R_{z(y)}(\delta)$ are rotations around axis $z$ $(y)$  with
Euler angle $\delta$.  
Using the Particle Data Group (PDG) parametrization for the phase space 
integrals \cite{PDG}, the  LO contribution to 
 $\im \overline A_{00+}$  
and $\im \overline B_{00+}$ of the 3-pion cut diagrams so read
\ba
\im \overline A_{00+}^{3\pi, LO}(s_1,s_2,s_3)&=&\fr{1}{16 (2\pi)^5}\int 
{\rm d} q_3^0 \ {\rm d} q_2^0 \ {\rm d} \alpha \ {\rm d} 
\cos\beta \ {\rm d} \gamma
\left[ A^{(p^2)}_{00+}(p_K,q_i) A_{3\pi,0}^{(p^2)}(q_i,p_i) \right] 
\ , \nn   \\
v_\pm \im \overline B_{00+}^{3\pi, LO}(s_1,s_2,s_3)&=&
\fr{1}{16 (2\pi)^5}\int 
{\rm d} q_3^0 \ {\rm d} q_2^0 \ {\rm d} \alpha \ {\rm d}\cos\beta 
\ {\rm d} \gamma
\left[ A^{(p^2)}_{++-}(p_K,q_i) A_{3\pi,\pm}^{(p^2)}(q_i,p_i) \right] \ ,
\nonumber\\ 
\label{eq:3p3pLO}
\ea
where  $ A_{3\pi,0}^{(p^2)}(q_i,p_i)$, $ A_{3\pi,\pm}^{(p^2)}(q_i,p_i)$ 
are the tree level amplitudes 
for $\pi^0\pi^0\pi^+\rightarrow \pi^0\pi^0\pi^+$
and $\pi^+\pi^+\pi^-\rightarrow \pi^0\pi^0\pi^+$. At this order  the 
are no contribution of these topologies to $\re \overline B_{00+}$.
We have computed  numerically the integrals in (\ref{eq:3p3pLO}). 
The correction induced  by (\ref{eq:3p3pLO}) to the other LO   
terms  is always much below the $0.5\%$.
We find so perfectly consistent to omit  these  corrections
 at this level of  precision.
Similar expressions and conclusions hold  for the FSI in the decay 
of $K_L\rightarrow \pi^0\pi^0\pi^0$. For this case one just has 
\ba
\im \overline A_{000}^{3\pi, LO}(s_1,s_2,s_3)&=&\fr{1}{16 (2\pi)^5}\int 
{\rm d} q_3^0 \ {\rm d} q_2^0 \ {\rm d} \alpha \ {\rm d} \cos\beta 
\ {\rm d} \gamma
\left[ A^{(p^2)}_{000}(p_K,q_i) \tilde A_{3\pi,0}^{(p^2)}(q_i,p_i) 
\right] \ , \nn   \\
v_\pm \im \overline B_{000}^{3\pi, LO}(s_1,s_2,s_3)&=&
\fr{1}{16 (2\pi)^5}\int 
{\rm d} q_3^0 \ {\rm d} q_2^0 \ {\rm d} \alpha \ {\rm d} \cos\beta 
\ {\rm d} \gamma
\left[ A^{(p^2)}_{+-0}(p_K,q_i)\tilde A_{3\pi,\pm}^{(p^2)}(q_i,p_i)
 \right] \ , \nonumber\\
\ea
where  $ \tilde A_{3\pi,0}^{(p^2)}(q_i,p_i)$, 
$\tilde A_{3\pi,\pm}^{(p^2)}(q_i,p_i)$ are the tree level amplitudes 
for $\pi^0\pi^0\pi^0\rightarrow \pi^0\pi^0\pi^0$
and $\pi^+\pi^-\pi^0\rightarrow \pi^0\pi^0\pi^0$.



\begin{thebibliography}{99}  
  
\bibitem{SOME}
 G. Ecker, hep-ph/0011026;  
A. Pich,  hep-ph/9806303.  
\bibitem{reviews}
G. Ecker,   
Prog.\ Part.\ Nucl.\ Phys.\  {\bf 35} (1995) 1;   
E. de Rafael,   
hep-ph/9502254;   
A. Pich,   
Rept.\ Prog.\ Phys.\  {\bf 58} (1995) 563.  
  
\bibitem{WEI79} S. Weinberg,   
Physica {\bf A 96} (1979) 327.  
  
\bibitem{GL84}  
J. Gasser and H. Leutwyler,  
Annals \ Phys.\  {\bf 158} (1984) 142;  
Nucl.\ Phys.\ {\bf B 250} (1985) 465.  
  
\bibitem{GPS03}  
 E.~G\'amiz, J.~Prades and I.~Scimemi,  
  J. High Energy Phys. {\bf 10} (2003) 042;  
hep-ph/0410150;  
hep-ph/0305164.  
  
\bibitem{SGP04}  
 I. Scimemi, E. G\'amiz and  J. Prades,  
Proc. of the $39^{th}$ Rencontres de Moriond on Electroweak Interactions   
and Unified Theories,  
p. 355, The Gioi Publishers (2005), hep-ph/0405204.  
  
\bibitem{PGS05}  
J. Prades, E. G\'amiz and I. Scimemi,  
Proc. of QCD'05, Montpellier, 4-8 July 2005,  
hep-ph/0509346.  
  
\bibitem{BKM91}  
V. Bernard, N. Kaiser and U.-G. Mei{\ss}ner,  
Nucl.\ Phys. {\bf B 357} (1991) 129.  
  
\bibitem{BDP02} J. Bijnens, P. Dhonte  and F. Persson,   
Nucl. Phys. {\bf B} 648 (2003) 317.  
  
\bibitem{KMW90}  
J. Kambor, J. Missimer and D. Wyler,  
Nucl.\ Phys. {\bf B 346} (1990) 17;  
Phys.\ Lett. {\bf B 261} (1991) 496.  
  
\bibitem{KDHMW92}   
J. Kambor, J.F. Donoghue, B.R. Holstein, J. Missimer and D. Wyler,  
 Phys.\ Rev.\ Lett.\  {\bf 68} (1992) 1818.  
  
\bibitem{BB04} J. Bijnens and F. Borg,   
Nucl. Phys.  {\bf B 697} (2004) 319;  
Eur. Phys. J. {\bf C 39} (2005) 347;  
A. Nehme, Phys.\ Rev.\ {\bf D 70} (2004) 094025.  
  
\bibitem{BB05} J. Bijnens and F. Borg,    
Eur.\ Phys.\ J. {\bf C 40} (2005) 383.  
  
\bibitem{CAB04} N. Cabibbo, Phys. Rev. Lett.  
{\bf 93} (2004) 121801.  
  
\bibitem{MMS97}
U.-G. Mei{\ss}ner, G. M\"uller and S. Steininger,
  Phys.\ Lett.\ B {\bf 406} (1997) 154
  [Erratum-ibid.\ B {\bf 407} (1997) 454];
  M. Knecht and R. Urech,
  Nucl.\ Phys.\ B {\bf 519} (1998) 329.

\bibitem{CI05} N. Cabibbo and G. Isidori,  
J. High Energy Phys. {\bf 03} (2005) 021.  
  
\bibitem{DIPP94}  
G. D'Ambrosio, G. Isidori, A. Pugliese and N. Paver,
  Phys.\ Rev.\ D {\bf 50} (1994) 5767
  [Erratum-ibid.\ D {\bf 51} (1995) 3975].

\bibitem{NA48} 
J.R. Batley {\it et al.}  [NA48/2 Collaboration],
Phys. Lett. {\bf B 633} (2006) 173;
S. Giudici [NA48/2 Collaboration], hep-ex/0505032.  
 
\bibitem{GAS06} J. Gasser, Talk at 
IV EURIDICE Collaboration Meeting,  Marseille 8-11 February 2006.

\bibitem{CGL00}  
G. Colangelo, J. Gasser and H. Leutwyler,  
  Phys. Lett. {\bf B 488} (2000) 261,  
Nucl.\ Phys. {\bf B 603} (2001) 125.  
  
\bibitem{PY04}  
J.R. Pel\'aez and F.J. Yndur\'ain,  
  Phys.\ Rev. {\bf D 71} (2005) 074016;  
 hep-ph/0412320.   
  
\bibitem{CGKR05}
 J. Gasser, private communication; 
G. Colangelo, J. Gasser, B. Kubis, and A. Rusetsky,
Phys. Lett. B {\bf 638} (2006) 187.

\bibitem{ANI03}
A.V. Anisovich, 
  Phys.\ Atom.\ Nucl.\  {\bf 66} (2003) 172
  [Yad.\ Fiz.\  {\bf 66} (2003) 175];
V.V. Anisovich and  A.A. Ansel'm, 
Usp. Fiz. Nauk. {\bf 88} (1966) 287
[Sov. Phys. Usp. {\bf 9} (1966) 117].

\bibitem{UE:05}
R. Unterdorfer and  G. Ecker,
J. High Energy Phys. {\bf 10} (2005) 017.

\bibitem{PDG}
S. Eidelman {\it et al.}  [Particle Data Group],
  Phys.\ Lett. {\bf B 592} (2004) 1.

\end{thebibliography}
\end{document}